\documentclass[11pt,aps,a4paper,eqsecnum,amsmath,amssymb
              ,superscriptaddress,notitlepage,nofootinbib%,endfloats*
              ,longbibliography
              ]{revtex4-1}
\pdfoutput=1

\usepackage{graphicx}
\usepackage{color}
\usepackage{tikz}
\usepackage{pgfplots}

\newcommand{\EQ}{\begin{equation}}
\newcommand{\EN}{\end{equation}}

\newcommand{\bear}{\begin{eqnarray}}
\newcommand{\ear}{\end{eqnarray}}

\newcommand{\ihalf}{\frac{i}{2}}

\begin{document}
\preprint{}

\title{The fine structure of the finite-size effects for the spectrum\\
  of the $OSp(n|2m)$ spin chain}

\author{Holger Frahm}
\affiliation{%
Institut f\"ur Theoretische Physik, Leibniz Universit\"at Hannover,
Appelstra\ss{}e 2, 30167 Hannover, Germany}

\author{M\'arcio J. Martins}
\affiliation{%
Departamento de F\'isica, Universidade Federal de S\~ao Carlos,
C.P. 676, 13565-905 S\~ao Carlos (SP), Brazil}

%\date{\today}
\begin{abstract}
  In this paper we investigate the finite-size properties of the spectrum of
  quantum spin chains with local spins taken to be the fundamental vector
  representation of the $OSp(n|2m)$ superalgebra.
\end{abstract}

%\pacs{05.20-y, 0.5.50+q, 04.20.Jb }

\maketitle

\section{Introduction}

Over the years exactly solvable one-dimensional quantum magnets have been
considered as suitable lattice regularization of two-dimensional space-time
models of quantum field theory. In principle the respective Bethe ansatz
solution offers us a non-perturbative framework to study the properties of the
spectrum of the respective spin chain Hamiltonian for large system sizes.  In
the case of a massless theory it has been showed that the finite size
corrections to the spectrum determine the conformal central charge and the
anomalous dimensions of the underlying conformal field theory
\cite{Cardy87}. The study of the finite-size effects of integrable spin chains
with generators on some simply laced Lie algebra $\mathrm{G}$ suggested that
their critical behaviour are governed by the properties of a field theory of
Wess-Zumino-Witten type on the same group $\mathrm{G}$.  By way of contrast
when the underlying invariance of the spin chain is based on supergroups the
identification of the respective field theory appears to be more involved
\cite{Saleur00}. Indeed, it has been observed that the finite size spectrum of
the $OSp(3|2)$ superspin chain present the unusual feature of having states
with the same conformal dimension as the trivial identity operator
\cite{MaNR98,FrMa15}.  Later on similar phenomena have been found to be
present in a staggered $sl(2|1)$ spin chain whose degrees of freedom alternate
between the fundamental and dual representations \cite{EsFS05} as well as in
staggered six-vertex model \cite{IkJS08}. The degeneracy of many states of the
spectrum was found to grow with the size of the chain and this was interpreted
as the signature of the existence of non-compact degrees of freedom in the
continuum limit \cite{IkJS08}.

The purpose of this paper is to study the subleading corrections to the
finite-size spectrum of a number of spin chains invariant by the $OSp(n|2m)$
super Lie algebra. The results obtained here extend in a substantial way our
recent analysis performed for the specific case of the $OSp(3|2)$ superalgebra
\cite{FrMa15}.  In particular, we find a tower of states over the lowest
energy with the same leading effective central charge $c_{\mathrm{eff}}$ as
the size of the chain ${L} \rightarrow \infty$.  More precisely,
denoting the eigenenergies of such set of states by $E_{k}(L)$ we have,
\begin{equation}
  E_{k}(L) - L e_{\infty} = \frac{\pi \xi c_{\mathrm{eff}}}{6L}
  +\frac{2\pi\xi}{L}\,\frac{\beta_{k}}{\log L}\,,
  \quad k=0,1,2,\cdots,k_{\infty}
\end{equation}
%where $k=0$ stands for suitable choice of a state in the tower of
%eigenenergies while 
where the integer $k_{\infty}$ is typically bounded by system size
$\mathrm{L}$. The symbol $e_{\infty}$ denotes the energy density of the ground
state in the thermodynamic limit while $\xi$ refers to the velocity of the
elementary low-lying excitations. We shall notice that the amplitude
$\beta_{k}$ can be connected to a subset of the possible eigenvalues
of the quadratic Casimir operator of the respective underlying $OSp(n|2m)$
superalgebra.

We recall here that the $OSp(n|2m)$ superspin chain realizes a gas of loops on
the square lattice in which intersections are allowed \cite{MaNR98}. The
integer $n$ and $m$ parameterize the fugacity $z$ given to every configuration
of closed loops which is $z=n-2m$. In the context of the loop model the above
peculiar finite-size behaviour was argued to be an indication that for $z <2$
the crossing of loops becomes a relevant perturbation driving the system to an
unusual critical phase \cite{JaRS03}. In particular it was conjectured that
the correlations functions in the loop model should be those of the Goldstone
phase of the $O(z)$ sigma model. The universal behaviour of the two point
correlators has long been computed in \cite{Polyakov75} and it was found to
decrease logarithmically with the distance.  More recently this calculation
has been extended to two point functions of operators composed by the product
of $k$ field components at the same point usually denominated $k$-leg
watermelon correlators \cite{NSSO13}.  This observable measures the
probability of $k$ distinct loop segments connecting two arbitrary lattice
points $x$ and $y$.  Here we shall argue that the asymptotic behaviour of such
correlation functions of the intersecting loop model can be inferred from the
finite-size amplitudes $\beta_{k}$ in analogy to the known connection among
critical exponents and finite-size scaling amplitudes \cite{Cardy87}.  More
precisely we observe that for large distances $r=|x-y|$ this
%We postulated that for large distances $r=|x-y|$ this
family of correlators can be rewritten as
\begin{equation}
  G_{k}(r) \sim 1/\ln(r)^{2(\beta_{k}-\beta_{k_0})}\,
\end{equation}
for a suitable choice of the $k_0$ state.

\section{The $OSp(n|2m)$ spin chain}

The vertex model with rational weights which is invariant by the superalgebra
$OSp(n|2m)$ was first discovered by Kulish in the context of the graded
formulation of the Yang-Baxter equation \cite{Kulish85}. The respective
$\mathrm{R}$-matrix $\mathrm{R}_{ab}(\lambda)$ with spectral parameter
$\lambda$ can be represented as a linear combination of three basic operators,
\begin{equation}
\label{RMA}
\mathrm{R}_{ab}(\lambda)= \lambda \mathrm{I}_{a} \otimes \mathrm{I}_{b} +\mathrm{P}_{ab} 
+ \frac{\lambda}{\frac{2-n+2m}{2}-\lambda} \mathrm{E}_{ab}
\end{equation}
where $\mathrm{R}_{ab}(\lambda)$ acts on the tensor product
$\mathrm{V}_a \times \mathrm{V}_b$ of two $(n+2m)$-dimensional graded
vector spaces and $\mathrm{I}_a$ denotes the identity matrix in one of
such spaces. The integers $n$ and $2m$ stand for the number of bosonic
(${b}$) and fermionic (${f}$) degrees of freedom.

The operator $\mathrm{P}_{ab}$ permutes two graded vector spaces and 
its expression is,
\begin{equation}
\mathrm{P}_{ab}= \sum_{i,j=1}^{n+2m} (-1)^{p_i p_j} 
e_{ij}^{(a)} \otimes 
e_{ji}^{(b)}
\end{equation}
where $p_i=0$ for the $n$ bosonic basis vectors while for the $2m$
fermionic coordinates we have $p_i=1$. The elementary matrices
$e_{ij}^{(a)} \in \mathrm{V}_a$ have only one non-vanishing element
with value 1 at row $i$ and column $j$.

The operator $\mathrm{E}_{ab}$ plays the role of a typical monoid
operator which can formally be represented as,
\begin{equation}
\mathrm{E}_{ab}= \sum_{i,j,l,k=1}^{n+2m} 
\alpha_{ij} \alpha^{-1}_{lk}e_{i l}^{(a)} \otimes 
e_{j k}^{(b)}
\end{equation}
where the non-null matrix elements $\alpha_{ij}$ are always $\pm 1$.  Their
precise distribution within the matrix $\alpha$ depends on the grading
sequence we set up for the basis of the vector space.  A convenient grading
sequence is the basis ordering $f_1 \cdots f_m {b}_1 \cdots {b}_n {f}_{m+1}
\cdots {f}_{2m}$ since it encodes in an explicit way the many ${U}(1)$
symmetries of the $OSp(n|2m)$ superalgebra.  For this choice of grading the
structure of the matrix $\alpha$ is \cite{MaRa97a},
\begin{equation}
\label{alpha}
\alpha=\left( \begin{array}{ccc} 
	O_{n \times m} &   O_{n \times m} & {\cal{I}}_{n \times n} \\
	O_{m \times m} &   {\cal{I}}_{m \times m} & O_{m \times n} \\
	-{\cal{I}}_{m \times m} &   O_{m \times m} & O_{m \times n} \\
	\end{array}
	\right)
\end{equation}
where $O_{ N \times N}$ and ${\cal{I}}_{ N \times N} $ are the null and the
anti-diagonal $N \times N$ matrices, respectively.  The matrix representation
for other grading choices can be obtained from Eq.(\ref{alpha}) by direct
permutation of the vector space basis.

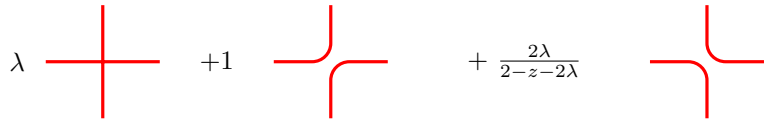
\begin{figure}[t]
\begin{center}

\begin{tikzpicture}[scale=1.5]
\draw (0.25,0) node {$\lambda$};
\draw (0.5,0) [color=red, very thick, rounded corners=7pt]
   +(0,0) -- +(1,0) +(0.5,0.5) -- +(0.5,-0.5);
\draw (2,0) node {$+1$}; 
\draw (2.5,0) [color=red, very thick, rounded corners=7pt]
  +(0,0) -- +(0.5,0) -- +(0.5,0.5)
  +(0.5,-0.5) -- +(0.5,0) -- +(1,0);
\draw (4.7,0) node {$+\,\,\frac{2\lambda}{2-z-2\lambda}$}; 
\draw (5.8,0) [color=red, very thick, rounded corners=7pt]
  +(0,0) -- +(0.5,0) -- +(0.5,-0.5)
  +(0.5,0.5) -- +(0.5,0) -- +(1,0);

\end{tikzpicture}
\end{center}
\caption{The local configurations contributing to the partition function of
  the intersecting loop model with fugacity $z=n-2m$ and their Boltzmann
  weights corresponding to the $R$-matrix (\ref{RMA}).\label{fig:bwloops}}
\end{figure}

In the intersecting loop model realized by this superspin chain the different
terms in the $R$-matrix (\ref{RMA}) correspond to the allowed local
configurations with Boltzmann weights given by the respective amplitudes, see
Figure~\ref{fig:bwloops}.
%%%%%%%%%%%%%%%%%%%%%%%%%%%%%%%%%%%%%%%%%%%%%%%%%%%%%%%%%%%%%%%%%%%%%%
%%%%%%%%%%%%%%%%%%%%%%%%%%%%%%%%%%%%%%%%%%%%%%%%%%%%%%%%%%%%%%%%%%%%%%
The Hamiltonian of the quantum $OSp(n|2m)$ spin chain is obtained by expanding
the transfer matrix of the respective vertex model at the special value of the
spectral parameter for which the $\mathrm{R}$ is proportional to the graded
permutator. Let us denote such transfer matrix by $T(\lambda)$ on a $L\times
L$ square lattice with toroidal boundary conditions.  It follows that this
operator can be written as the supertrace of an auxiliary operator called
monodromy matrix \cite{Kulish85},
\begin{equation}
T(\lambda)= \sum_{i=1}^{n+2m} (-1)^{p_i} {\cal{T}}_{ii}(\lambda)
\end{equation}
where elements of the monodromy matrix ${\cal{T}}_{ij}$ are given by
an ordered product of $\mathrm{R}$-matrices acting on the same
auxiliary space but with distinct quantum space components,
\begin{equation}
  {\cal{T}}(\lambda)= \mathrm{R}_{0L}(\lambda) \mathrm{R}_{0L-1}(\lambda) \dots
  \mathrm{R}_{01} (\lambda) 
\end{equation}

As usual considering the logarithmic derivatives of $T(\lambda)$
around the regular point $\lambda=0$ we obtain the local integrals of
motion. The first non-trivial charge turns to be the Hamiltonian whose
expression is,
\begin{equation}
\label{HAM}
\mathrm{H}= \epsilon \sum_{i=1}^{L}
\left[\mathrm{P}_{i,i+1}+\frac{2}{2-n+2m}\mathrm{E}_{i,i+1} \right] \,,
\end{equation}
where periodic boundary conditions for both bosonic and fermionic
degrees of freedom is assumed. The anti-ferromagnetic regime for $n-2m
<2$ requires the choice $\epsilon=-1$ while for $n-2m >2$ we need to
take $\epsilon=+1$.

The spectrum of this Hamiltonian can be studied by Bethe ansatz methods and is
parametrized by solutions to a set of algebraic Bethe equations. Since these
Bethe equations depend on the particular choice of the grading their root
configurations are grading dependent.  We can however infer on the infinite
volume properties of such superspin chain without the need of choosing an
specific Bethe ansatz solution \cite{MaNR98}.  This can be done establishing
certain a functional relation for the largest eigenvalue of the transfer
matrix usually by means of the matrix inversion method \cite{Stro79,Baxt82a}.
In our case this identity can be derived combining the unitarity property of
the $\mathrm{R}$-matrix (\ref{RMA}) together with its crossing symmetry under
translation $\lambda \rightarrow (2-n+2m)/2 -\lambda$ of the spectral
parameter. Let us denote by $[\Lambda_0(\lambda)]^{L}$ the largest eigenvalue
which dominates the partition function of the vertex model per site in the
thermodynamic limit.  We find that $\Lambda_0(\lambda)$ satisfies the
following constraint,
\begin{equation}
  \label{FUN}
  \Lambda_0(\lambda) \Lambda_0(\lambda+\frac{2-n+2m}{2}) =\frac{(\lambda^2-1)(
    \lambda+\frac{2-n+2m}{2})}{\lambda}
\end{equation}

Using unitarity $\Lambda_0(\lambda) \Lambda_0(-\lambda)=(1-\lambda^2)$ we can
solve the above functional relation under the assumption of analyticity in the
region $0 \leq \lambda <|2-n+2m|/2$. The final result is,
\begin{eqnarray}
  \Lambda_0(\lambda)&=&\left[\frac{(2-n+2m)^2}{\frac{|2-n+2m|}{2}-\lambda} \right] 
  \frac{\Gamma\left(1+\frac{\lambda}{|2-n+2m|}\right)
    \Gamma\left(\frac{1}{2}+\frac{1}{|2-n+2m|}+\frac{\lambda}{|2-n+2m|}\right)
    \Gamma\left(\frac{3}{2}-\frac{\lambda}{|2-n+2m|}\right)}
  {\Gamma\left(\frac{1}{2} +\frac{\lambda}{|2-n+2m|}\right)
    \Gamma\left(\frac{1}{|2-n+2m|} +\frac{\lambda}{|2-n+2m|}\right)
    \Gamma\left(1-\frac{\lambda}{|2-n+2m|}\right)} \nonumber \\
  &\times& \frac{\Gamma\left(1+\frac{1}{|2-n+2m|}-\frac{\lambda}{|2-n+2m|}\right)}
  {\Gamma\left(\frac{1}{2}+\frac{1}{|2-n+2m|}-\frac{\lambda}{|2-n+2m|}\right)}
\end{eqnarray}
where $\Gamma(x)$ is the Euler's integral of the second kind.

The ground state energy per site $e_{\infty}$ of the $OSp(n|2m)$ spin
chain (\ref{HAM}) is obtained by taking the logarithmic derivative of
$\Lambda_0(\lambda)$ at the spectral point $\lambda=0$. After some
simplifications we find,
\begin{equation}
e_{\infty}= -\frac{2}{|2-n+2m|} \left[ \psi \left(\frac{1}{2} +\frac{1}{|2-n+2m|} \right)
-\psi \left(\frac{1}{|2-n+2m|} 
\right) +2\ln(2) \right]+1
\end{equation}
where $\psi(x)=\frac{d \ln \Gamma(x)}{dx}$ is the Euler $\mathrm{psi}$ function.

The same reasoning as above can be used to obtain the dispersion
relation for the low-lying excitations, see for instance
Ref.~\cite{Klumper90}. These states correspond to next largest
eigenvalues of the transfer matrix and their ratios with the ground
state $[\Lambda_0(\lambda)]^{L}$ defines the excitation function
$\gamma(\lambda)$. Considering that Eq.(\ref{FUN}) applies also for
the excitations such function is expected to satisfies the constraint
$\gamma(\lambda) \gamma(\lambda+\frac{2-n+2m}{2}) =1$.  This means
that $\gamma(\lambda)$ has the real period $|2-n+2m|$ and consequently
it can be expressed in terms of product of trigonometric functions. We
can now follow the reasoning discussed in \cite{Klumper90} and
conclude that the dispersion relation $e(p)$ for the low-lying
excitations with momenta $p$ is,
\begin{equation}
e(p) =\frac{2\pi}{|2-n+2m|} \sin(p)
\end{equation}
and therefore the speed of sound is $\xi =\frac{2\pi}{|2-n+2m|}$. 

We would like to note that for the results so far it has implicitly been
assumed that $n-2m \neq 2$.  For $n-2m=2$ we can not derive the Hamiltonian
from the $\mathrm{R}$-matrix (\ref{RMA}) since there is no point $\lambda_0$
such that $\mathrm{R}_{ab}(\lambda_0) \sim \mathrm{P}_{ab}$.  These are the
cases in which the Killing form of the $OSp(n|2m)$ superalgebra is
degenerated.  One way to circumvent this problem is to scale the spectral
parameter $\lambda \rightarrow \lambda (2-n+2m)/2$ and afterwards take the
limit $n-2m \rightarrow 2$ in Eq.(\ref{RMA}) to obtain,
\begin{equation}
\label{RMA1}
\mathrm{\overline{R}}_{ab}(\lambda)= \mathrm{P}_{ab} 
+ \frac{\lambda}{1-\lambda} \mathrm{E}_{ab}~~~\mathrm{for}~~~n-2m=2
\end{equation}
which is the $\mathrm{R}$-matrix of the so-called Temperley-Lieb model
with $E_{ab}^2=2 E_{ab}$ \cite{Baxt82a}. We note that in this case the
respective loop model realization does not permit configurations
involving intersecting paths since the identity operator is not
present in the $\mathrm{R}$-matrix (\ref{RMA1}).  We further recall
that for $n=2$ and $m=0$ the vertex model corresponds to the isotropic
six-vertex model.
The expression of the respective anti-ferromagnetic Hamiltonian is,
\begin{equation}
\label{HAM1}
\mathrm{\overline{H}}=-\sum_{i=1}^{L} \mathrm{E}_{i,i+1}
~~~\mathrm{for}~~~n-2m=2
\end{equation}

The inversion method can also provide us with exact results for the vertex model
with weights based on the $\mathrm{R}$-matrix (\ref{RMA1}). It turns out that the
respective partition function per site is,
\begin{equation}
\overline{\Lambda_0}(\lambda)= \frac{2}{1-\lambda} \frac{\Gamma\left(1+\frac{\lambda}{2} \right)
\Gamma\left(\frac{3}{2}-\frac{\lambda}{2} \right)}
{\Gamma\left(1-\frac{\lambda}{2} \right)
\Gamma\left(\frac{1}{2}+\frac{\lambda}{2} \right)}
~~~\mathrm{for}~~~n-2m=2
\end{equation}
while the ground state energy and dispersion relation associated 
to the Hamiltonian (\ref{HAM1}) are,
\begin{equation}
\bar{e}_{\infty}= -2 \ln(2)~~~\mathrm{and}~~~\bar{e}(p)=\pi \sin(p)
~~~\mathrm{for}~~~n-2m=2
\end{equation}

From the above results we conclude that the bulk behaviour depends only
on the loop model fugacity $z=n-2m$. In next sections we shall present
evidences that this feature still remains valid for the central charge
and for the compact part of the critical exponents of the underlying
conformal field theory.

\section{Small size results}

In order to gain some insight on the spectrum properties 
of the $OSp(n|2m)$ superspin chain we have numerically 
diagonalize the 
respective Hamiltonians for lattice sizes $L \leq 8$.
We have limited our analysis to Hamiltonians with 
maximum number of seven
states per site $n+2m=7$. We find that the ground state is 
generically degenerated for spin chains with 
$n-2m \leq 0$ while when $n-2m \geq 1$ the ground state is
always a singlet for $L$ even.
In Table \ref{tab1} we present the ground state 
degeneracies for the $OSp(n|2m)$
spin chains studied in this paper for even and odd lattice sizes.
\begin{table}[ht]
\begin{center}
\begin{tabular}{|c||c|c|}
  \hline
& even~L & odd~L   \\ \hline \hline
$OSp(1|2)$  & 3  & 3    \\ \hline
$OSp(3|4)$ & 23  & 7     \\ \hline 
$OSp(2|2)$ & 8   & 4     \\ \hline
$OSp(3|2)$ & 1   & 5    \\ \hline
$OSp(2|4)$ & 16  & 32      \\ \hline
$OSp(5|2)$ & 1 & 7    \\ \hline
\end{tabular}
\caption{Ground state energy degeneracies for even and odd lattice sizes.}
\label{tab1}
\end{center}
\end{table}

We have noted that for a fixed fugacity $n-2m$ the eigenspectrum are
basically the same apart degeneracies up to the size $L=4$ for
distinct values of $n$ and $m$. Considerable number of new eigenvalues
start to emerge for $L=6$ but they occur at the higher energy part of
the spectrum.  These findings suggests that for large enough $L$ the
spectrum should satisfies the following sequence of inclusions,
\begin{equation}
  \label{eq:specincl}
  \mathrm{spec}[OSp(n|2m)] \subset \mathrm{spec}[OSp(n+2|2m+2)]  
  \subset \mathrm{spec}[OSp(n+4|2m+4)] \subset \dots
\end{equation}
such that the ground state and the low-lying excitations for a given
fugacity $n-2m$ is described by the superspin chain with the lowest
possible values of the integers $n$ and $m$. 
This feature is present in spin chains with different supergroup symmetries,
e.g.\ for $gl(m|n)$ where a spectral embedding of models with given $m-n$ has
been observed \cite{Cand11}.

This above observation can be used in order to predict the value for the
effective central charge. For $n-2m \geq 2$ the sequence can be started with
the orthogonal invariance $O(n-2m)$ and the respective conformal field theory
should be that of the Wess-Zumino-Witten model on this group see for instance
\cite{Martins90a,Martins91}. The partition function is expected to be
dominated by $n-2m$ Ising degrees of freedom and therefore the central charge
is,
\begin{equation}
  c_{\mathrm{eff}}=(n-2m)/2~~\mathrm{for}~~n-2m \geq 2
\end{equation}

On the other hand when $n-2m < 2$ the orthogonal invariance is somehow broken
and the partition function is effectively dominated by $n-2m-1$ bosonic
degrees of freedom with effective central charge \cite{MaNR98},
\begin{equation}
  \label{osp-ceff}
  c_{\mathrm{eff}}=n-2m-1\quad\mathrm{for}\quad n-2m <2\,.
\end{equation}

At this point we remark that in the context of the intersecting loop model
these two regimes are distinguished by the behaviour of the respective
Boltzmann weights. We note that for $n-2m < 2$ the three weights in
Eq.(\ref{RMA}) can be chosen positive and consequently they can be interpreted
as probabilities. However when $n-2m >2$ one of the weights is always negative
and the probability interpretation is lost. Therefore it is not a surprise
that the continuum limit of these regimes are described by two different
conformal field theories.

In next section we shall begin our study of the finite-size effects for large
$L$ for the superspin chains in Table \ref{tab1} by using convenient grading
choice for the Bethe ansatz solution. We will investigate two specific
sequences of models with the same fugacity and argue that the potential extra
eigenvalues does not lead to new conformal dimensions.

%\end{document}

%%%%%%%%%%%%%%%%%%%%%%%%%%%%%%%%%%%%%%%%%%%%%%%%%%%%%%%%%%%%%%%%%%%%%%
\section{Finite size effects}
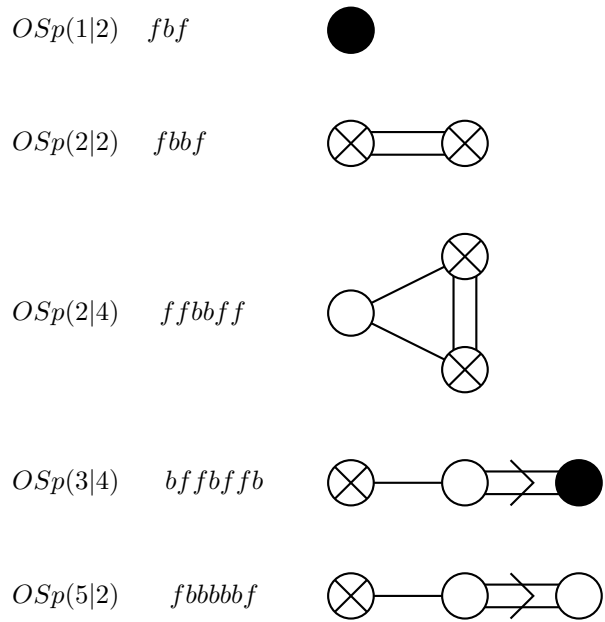
\begin{figure}[th]
\begin{tikzpicture}[thick,scale=1.5]

        ! black circle @ (1.5,0)
        \draw[fill=black] (1.5,0) circle(.2);

        \node at (-1,0) {$OSp(1|2)$};

        \node at (-0.1,0) {${fbf}$};

% \end{tikzpicture}

% \vspace{.4in}

% \begin{tikzpicture}[thick,scale=1.5]

\pgftransformyshift{-1cm}

        \draw (1.5,.1) -- (2.5,.1);
        \draw (1.5,-.1) -- (2.5,-.1);

        ! crossed circle @ (1,0)
        \draw[fill=white] (1.5,0) circle(.2);
        \draw (1.5,0) +( 45:.2) -- +(225:.2);
        \draw (1.5,0) +(135:.2) -- +(315:.2);

        ! crossed circle @ (2.5,0)
        \draw[fill=white] (2.5,0) circle(.2);
        \draw (2.5,0) +( 45:.2) -- +(225:.2);
        \draw (2.5,0) +(135:.2) -- +(315:.2);

        \node at (-1,0) {$OSp(2|2)$};

        \node at (0,0) {${fbbf}$};

% \end{tikzpicture}

% \vspace{.4in}

% \begin{tikzpicture}[thick,scale=1.5]
\pgftransformyshift{-1.5cm}

        \draw (1.5,0) -- (2.5,0.5);

        \draw (1.5,0) -- (2.5,-0.5);

        \draw (2.4,0.5) -- (2.4,-0.5);
        \draw (2.6,0.5) -- (2.6,-0.5);

        \draw[fill=white] (1.5,0) circle(.2);

        \draw[fill=white] (2.5,0.5) circle(.2);
        \draw (2.5,0.5) +( 45:.2) -- +(225:.2);
        \draw (2.5,0.5) +(135:.2) -- +(315:.2);

        \draw[fill=white] (2.5,-0.5) circle(.2);
        \draw (2.5,-0.5) +( 45:.2) -- +(225:.2);
        \draw (2.5,-0.5) +(135:.2) -- +(315:.2);

        \node at (-1,0) {$OSp(2|4)$};

        \node at (0.2,0) {${ffbbff}$};

% \end{tikzpicture}

% \vspace{.4in}

% \begin{tikzpicture}[thick,scale=1.5]
\pgftransformyshift{-1.5cm}

        \draw (1.5,0) -- (2.5,0);

        \draw (2.5,.1) -- (3.5,.1);
        \draw (2.5,-.1) -- (3.5,-.1);
%       \draw (1.4,0.2) -- (1.6,0) -- (1.4,-0.2);
        \draw (2.9,0.2) -- (3.1,0) -- (2.9,-0.2);
        
        \draw[fill=white] (1.5,0) circle(.2);
       \draw (1.5,0) +( 45:.2) -- +(225:.2);
       \draw (1.5,0) +(135:.2) -- +(315:.2);

        \draw[fill=white] (2.5,0) circle(.2);
  %     \draw (1,0) +( 45:.2) -- +(225:.2);
  %     \draw (1,0) +(135:.2) -- +(315:.2);

        \draw[fill=black] (3.5,0) circle(.2);
        
        % ! black circle @ (3,0)
        % \draw[fill=black] (3,0) circle(.2);
        
        \node at (-1,0) {$OSp(3|4)$};
        \node at (0.3,0) {${bffbffb}$};
         
% \end{tikzpicture}

% \vspace{.4in}

% \begin{tikzpicture}[thick,scale=1.5]
\pgftransformyshift{-1cm}

        \draw (1.5,0) -- (2.5,0);

        \draw (2.5,.1) -- (3.5,.1);
        \draw (2.5,-.1) -- (3.5,-.1);
        \draw (2.9,0.2) -- (3.1,0) -- (2.9,-0.2);
        
        ! crossed circle @ (1.5,0)
        \draw[fill=white] (1.5,0) circle(.2);
        \draw (1.5,0) +( 45:.2) -- +(225:.2);
        \draw (1.5,0) +(135:.2) -- +(315:.2);

        ! white circle @ (2.5,0)
        \draw[fill=white] (2.5,0) circle(.2);

        ! white circle @ (3.5,0)
        \draw[fill=white] (3.5,0) circle(.2);

        \node at (-1,0) {$OSp(5|2)$};

        \node at (0.3,0) {${fbbbbbf}$};
         
\end{tikzpicture}

  \caption{The Dynkin diagram with the respective basis ordering for the
    superalgebras studied in this paper. The bosonic roots are represented by
    a white dot while the fermionic ones by a black dot or a crossed
    dot.\label{fig:dynkin}}
\end{figure}

In this section we will investigate the finite size properties of the super
spin chains with the help of their Bethe ansatz solution.  As mentioned above
it is a common feature of integrable spin models based on super Lie algebras
that the Bethe equations for the rapidities parametrizing their spectrum
depend on the choice of grading.
In a first step we have to choose the formulation which is most
convenient for the numerical solution of the respective Bethe ansatz
equations for large system sizes.  By now it is well know that for
rational vertex models there exists a direct connection between the
form of the Bethe ansatz equations with the specific Dynkin diagram
representation of the underlying superalgebra.  In Figure
\ref{fig:dynkin} we exhibit the diagrams with the respective grading
ordering for the orthosympletic superalgebras suitable for each super
spin chain studied in this paper.
The explicit form of the Bethe equations and the basic root
distributions is presented in the next subsections.

Based on the numerical solution of the Bethe equations we can analyze
the finite size scaling of the spectrum.  For a conformally invariant
theory the finite size gaps are expected to scale as \cite{BlCN86,Affl86}
\begin{equation}
  X_{\mathrm{eff}}(k;L) = \frac{L}{2\pi\xi}\left(E_{k}(L) -L
  \epsilon_\infty\right) \to X_{k}-\frac{c_{\mathrm{eff}}}{12}\,. 
\end{equation}
where $X_k$ are the scaling dimensions of the corresponding operator
in the continuum limit and the effective central charge
$c_{\mathrm{eff}}$ governs the finite size scaling of the ground state
energy $E_0(L)$ of the lattice model.  Similarly, from the momentum of
the states the conformal spin of the corresponding operator can be
determined, $s(k;L) = (L/2\pi)(P_k(L) - \mathcal{P}_0)$.

As we shall see below the spectrum of scaling dimensions of the $OSp(n|2m)$
models is highly degenerate in the thermodynamic limit.  In a finite system
this degeneracy is lifted by subleading corrections to scaling which can be
studied in conformal perturbation theory \cite{Card86a,Card86c}.  With respect
to the conformally invariant fixed point the lattice Hamiltonian of the
isotropic $OSp(n|2m)$ superspin chains is perturbed by a marginally irrelevant
operator.  If the coupling constant $g$ is initially small the effective
coupling at scale $L$ vanishes as $g(L)\sim 1/\log L$ and the corrections to
scaling take the universal form
\begin{equation}
  X(k;L) \simeq X_{k} + \frac{\beta(k)}{\log L}\,. 
\end{equation}
This logarithmic dependence on the system size requires information on the
spectrum for large system sizes to reliably determine the scaling dimensions.
Below we shall use this prediction to determine both them and the amplitudes
of $\beta(k)$ extrapolating finite size date for lattice systems with up to
several thousand sites based on the assumption that the corrections to scaling
are rational functions of $1/\log L$.

% \subsection{$n-2m=-3$}
% $OSp(1|4)$
%----------------------------------------------------------------------
\subsection{$n-2m=-2$: $OSp(2|4)$}
For the $OSp(2|4)$ model it turns out to be most convenient to use the
grading Bethe equations for the grading $ffbbff$
\begin{equation}
  \label{bae-o24}
\begin{aligned}
  &\left(\frac{\lambda_j^{(1)}+\ihalf}{\lambda_j^{(1)}-\ihalf}\right)^L
  = \prod_{k =1, k\ne j}^{N_1}
  \frac{\lambda_j^{(1)}-\lambda_k^{(1)}+i}{\lambda_j^{(1)}-\lambda_k^{(1)}-i}\,
  \prod_{k =1}^{N_+}
  \frac{\lambda_j^{(1)}-\lambda_k^{(+)}-\ihalf}{
        \lambda_j^{(1)}-\lambda_k^{(+)}+\ihalf}\,
  \prod_{k =1}^{N_-}
  \frac{\lambda_j^{(1)}-\lambda_k^{(-)}-\ihalf}{
        \lambda_j^{(1)}-\lambda_k^{(-)}+\ihalf}\,,
  \quad j=1\ldots N_1\,,\\
  & \prod_{k =1}^{N_1}
  \frac{\lambda_j^{(+)}-\lambda_k^{(1)}+\ihalf}{
        \lambda_j^{(+)}-\lambda_k^{(1)}-\ihalf}
  = \prod_{k =1}^{N_-}
  \frac{\lambda_j^{(+)}-\lambda_k^{(-)}+ i}{
        \lambda_j^{(+)}-\lambda_k^{(-)}- i}\,,\quad j=1\ldots N_+\,,\\
  & \prod_{k =1}^{N_1}
  \frac{\lambda_j^{(-)}-\lambda_k^{(1)}+\ihalf}{
        \lambda_j^{(-)}-\lambda_k^{(1)}-\ihalf}
  = \prod_{k =1}^{N_+}
  \frac{\lambda_j^{(-)}-\lambda_k^{(+)}+ i}{
        \lambda_j^{(-)}-\lambda_k^{(+)}- i}\,,\quad j=1\ldots N_-\,.
  \end{aligned}
\end{equation}
The number of Bethe roots on the three levels determine the eigenvalues of the
conserved $U(1)$ charges from the Cartan subalgebra of $OSp(2|4)$.  The energy
of the state parametrized by a solution $\{\lambda_k^{(1)}\} \cup
\{\lambda_k^{(+)}\} \cup \{\lambda_k^{(-)}\}$ to these equations is
\begin{equation}
\label{ene-o24}
  E = L - \sum_{k=1}^{N_1} \frac{1}{\left(\lambda_k^{(1)}\right)^2+\frac14}\,.
\end{equation}

The roots of (\ref{bae-o24}) corresponding to the ground state and many of the
low-lying excitations are found to be real with finite densities $N_1/L\to1$,
$N_\pm/L\to\frac12$ in the thermodynamic limit.  This fact allows to study
their finite size scaling analytically based on linear integral equations
\cite{Sogo84,VeWo85,BoIK86}.  In the present case we find that the Bethe
ansatz integral equations have a singular kernel, similar as in the staggered
$sl(2|1)$ superspin chains and the staggered six-vertex model where this has
been found to lead to a continuous spectrum of scaling dimensions
\cite{EsFS05,FrMa11,FrMa12,IkJS08}.  Labelling the charge sectors of the model
by the quantum numbers $n_1=L-N_1$, $n_2=L-N_+-N_-$ and $n_3=N_+-N_-$ and the
corresponding vorticities $m_{k=1,2,3}$ the resulting scaling dimensions of
primary fields are
\begin{equation}
  \label{xdims-o24}
\begin{aligned}
  X^{(2|4)}_{\mathrm{eff}} (n_k,m_k;L) \to 
    \frac14\left(n_1^2 + (n_1-n_2)^2 + {\epsilon}\,n_3^2
    \right)
  + \frac12\left( m_1^2 + (m_1+2m_2)^2 +
      \frac{1}{\epsilon}m_3^2\right) %\right) 
    - \frac14\,
  \end{aligned}
\end{equation}
and their conformal spin is $s(n_k,m_k) = \sum_k n_km_k$.
To derive (\ref{xdims-o24}) we have introduced the small parameter $\epsilon$
to regularize the singularity of the kernel.  By construction the quantum
numbers $n_k$ are integers while the vorticities take integer or half-odd
integer values according to the selection rules
\begin{equation}
  m_1 \sim \frac12\,n_2\bmod 1\,,\quad
  m_2 \sim \frac12\left(n_1- n_2 +1 \right)\bmod 1\,.
\end{equation}
$m_3$ is always integer.  In the limit $\epsilon \to 0^+$ scaling dimensions
with the same $n_3$ become degenerate while the vorticity $m_3$ is constraint
to be $0$ for states in the low energy spectrum (i.e.\ for operators with
finite scaling dimension $X$).  Taking these constraints into account we find
that the conformal weights in the low energy effective theory are non-negative
integers.

The ground state of the chain for even $L$ appears in the sector with
$(n_1,n_2,n_3)=(1,2,0)$ (and also $(1,0,0)$) and $m_1=m_2=0$. Both from
(\ref{xdims-o24}) and the extrapolation of data obtained by numerically
solving (\ref{bae-o24}) we find
\begin{equation}
  X_{\mathrm{eff}}^{(2|4)}\left(\mathbf{n}=(1,2,0);L\right) 
  \to \frac12-\frac14 = -\frac{c_{\mathrm{eff}}}{12}
\end{equation}
with the effective central charge $c_{\mathrm{eff}}=-3$, in agreement
with (\ref{osp-ceff}).

\begin{figure}[t]
  \includegraphics[width=0.7\textwidth]{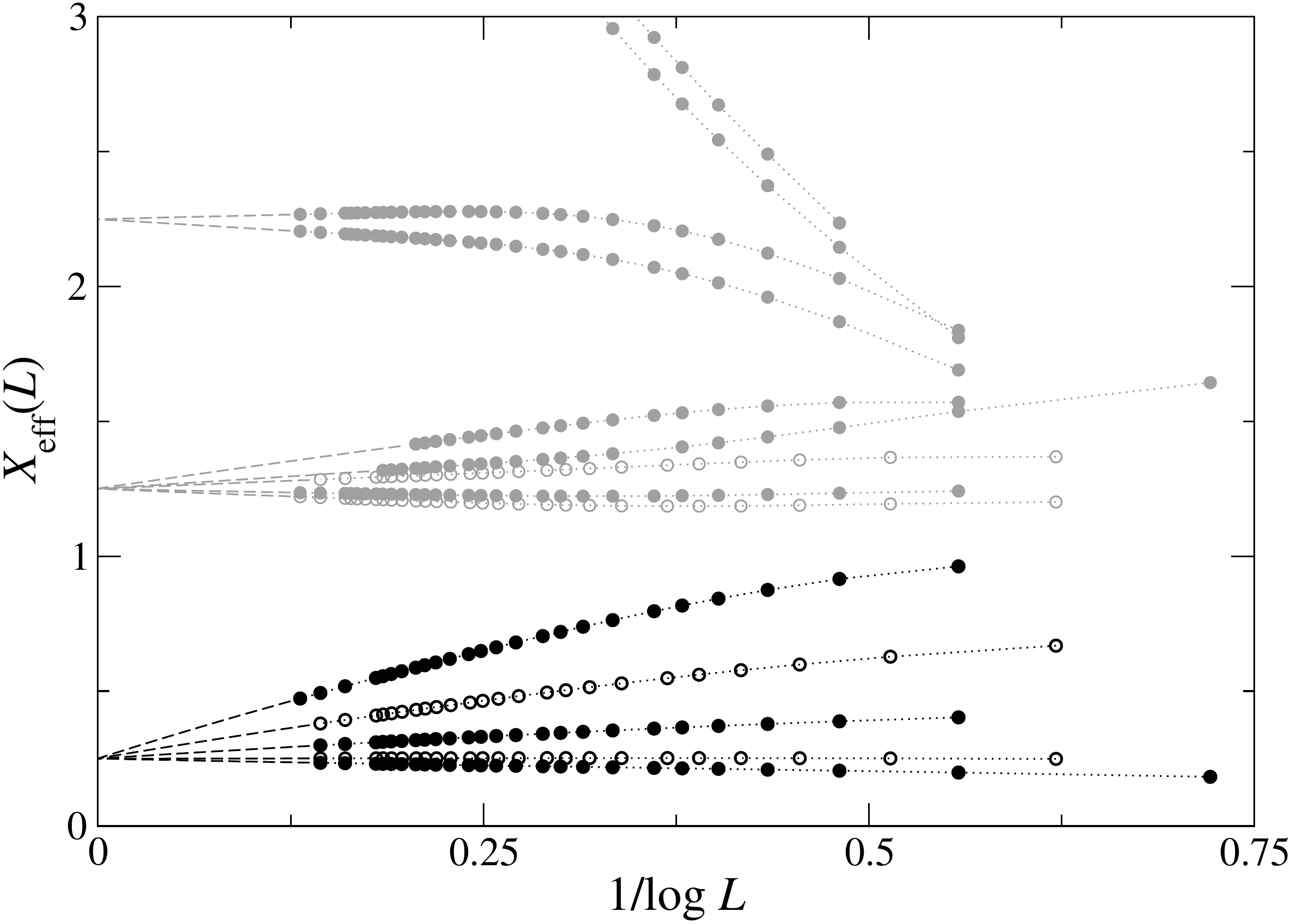}
  \caption{Finite size spectrum of the $OSp(2|4)$ superspin chain.  Displayed
    are the effective scaling dimensions $X_{\mathrm{eff}}(L)$ vs.\ $1/\log
    L$.  
    Black symbols denote levels from the lowest tower (\ref{o24_subl}) of
    scaling dimensions in the sectors $(1,2,k)$, $k=0,1,\ldots,4$, filled (open)
    symbols are data from the solution of the Bethe equations for chains of
    even (odd) length.  Grey symbols are higher excitations.  Dashed lines are
    extrapolations based on a rational dependence on $1/\log L$.
    \label{fig:fso24} }
\end{figure}

The lowest excitations in the sectors $(n_1,n_2,n_3)=(1,2,k)$ with
$|k|=1,2,3,\ldots\sim L\bmod 2$.  Among these is the lowest energy state of
the odd length super spin chains for $k=\pm1$ -- are also described by real
Bethe roots.  As expected from (\ref{xdims-o24}) they exhibit the same leading
finite size scaling as the ground state but different subleading corrections,
see Figure~\ref{fig:fso24}.
From our numerical data based on the solution of the Bethe equations we find
that the corrections to the scaling dimension of these states vanish as
$1/\log L$, as expected from perturbative renormalization group analysis of
the low temperature Goldstone phase of the loop models with $n-2m<2$
\cite{JaRS03}.  Analyzing the subleading corrections in detail we find
\begin{equation}
\label{o24_subl}
  X^{(2|4)}({(1,2,k)};L) \simeq \frac{\beta^{(2|4)}(k)}{\log L}\,,
    \quad
    \beta^{(2|4)}(k) = \frac{(k+1)(k-1)}{8}\,.
\end{equation}

Similar groups of excitations parameterized by real Bethe roots appear in the
sectors $(n_1,n_2,n_3)=(2,3,k)$ and $(2,4,k)$ with $n_2+k\sim L\bmod 2$, see
Figure~\ref{fig:fso24}.  The finite size analysis shows that the corresponding
primaries have a scaling dimension $X_{(2,3,k)}=1$ and $X_{(2,4,k)}=2$, their
conformal spins are $s=1$ and $s=0,2$, respectively.  Again, the subleading
corrections to finite size scaling are found to vanish as $1/\log L$ with
$k$-dependent amplitudes.

Among the remaining low energy levels in the spectrum of small systems found
by exact diagonalization we have identified (see Figure~\ref{fig:fso24})
\begin{itemize}
\item a descendent of the ground state with $X=1$, $s=1$ in the
  $(n_1,n_2,n_3)=(1,2,0)$ sector described by a Bethe root configuration
  containing a single $2$-string of complex conjugate Bethe roots
  $\lambda_{0\pm}^{(1)}\simeq \lambda_0\pm i/2$ with real $\lambda_0$ in
  addition to the real ones.
\item two states in the sectors $(2,4,0)$ and $(2,4,2)$ disappear from the low
  energy spectrum as the system size is increased.  Such behaviour is expected
  for levels violating the constraint $m_3=0$.
\end{itemize}

% ----------------------------------------------------------------------
\subsection{$n-2m=-1$}
For the $OSp(1|2)$ model the Bethe equations are \cite{Kulish85,Mart95}
\begin{equation}
  \label{bae-o12}
  \left(\frac{\lambda_j+\ihalf}{\lambda_j-\ihalf}\right)^L
  = \prod_{k =1}^{L-2n}
  \frac{\lambda_j-\lambda_k+i}{\lambda_j-\lambda_k-i}\,
  \frac{\lambda_j-\lambda_k-\ihalf}{\lambda_j-\lambda_k+\ihalf}\,,
  \quad j=1\ldots L-2n\,.
\end{equation}
Solutions of these equations parametrize highest weight states for
$(4n+1)$-dimensional irreducible representations of $OSp(1|2)$ with
superspin $J=n$, $n=0,\frac12,1,\frac32,\ldots$ and energy
\begin{equation}
  \label{ene-o12}
  E = L - \sum_{j=1}^{L-2n} \frac1{\lambda_j^2+\frac14}\,.
\end{equation}
The operator content of the effective theory describing this superspin chain
at low energies is known from Ref.~\cite{Mart95}, the primary fields have
scaling dimensions
\begin{equation}
  X_{\mathrm{eff}}^{(1|2)}(n,m;L) \to n^2+m^2 - \frac{1}{12}
\end{equation}
for states with superspin $J=n$ and vorticity $m$ subject to the constraint
$(n+m)\in\mathbb{Z}+\frac12$.\footnote{Note that there exist highly excited
  states for which this constraint is violated.}  Hence, for the triplet
ground state $(n,m)=(\frac12,0)$ we find the central charge
$c_{\mathrm{eff}}=-2$.
The finite size scaling and finite temperature properties of the $OSp(1|2)$
superspin has recently been studied based on a formulation of the Bethe ansatz
in terms of nonlinear integral equations \cite{TaRi17}.  As a consequence of
the boundary conditions used in that work the effective central charge
obtained from the low temperature behaviour differs from the one appearing in
the finite size scaling behaviour of the ground state.
We note that the ground state is degenerate (up to subleading corrections to
scaling) with the lowest singlet, $(n,m)=(0,\frac12)$.

As discussed above, the spectrum of the $OSp(1|2)$ superspin chain is a subset
of that of the $OSp(3|4)$ model.  Therefore we discuss the finite size scaling
in the context of the latter based on the Bethe ansatz for grading $bffbffb$ 
%----------------------------------------------------------------------
%
\begin{equation}
  \label{bae-o34}
\begin{aligned}
  &\left(\frac{\lambda_j^{(1)}+\ihalf}{\lambda_j^{(1)}-\ihalf}\right)^L
  = \prod_{k =1}^{N_2}
  \frac{\lambda_j^{(1)}-\lambda_k^{(2)}+\ihalf}{
        \lambda_j^{(1)}-\lambda_k^{(2)}-\ihalf}\,,
  \quad j=1\ldots N_1\,,\\
  & \prod_{k =1}^{N_1}
  \frac{\lambda_j^{(2)}-\lambda_k^{(1)}+\ihalf}{
        \lambda_j^{(2)}-\lambda_k^{(1)}-\ihalf}
  = \prod_{k =1, k\ne j}^{N_2}
  \frac{\lambda_j^{(2)}-\lambda_k^{(2)}+ i}{
        \lambda_j^{(2)}-\lambda_k^{(2)}- i}\,
  \prod_{k =1}^{N_3}
  \frac{\lambda_j^{(2)}-\lambda_k^{(3)}-\ihalf}{
        \lambda_j^{(2)}-\lambda_k^{(3)}+\ihalf}\,,\quad j=1\ldots N_2\,,\\
  & \prod_{k =1}^{N_2}
  \frac{\lambda_j^{(3)}-\lambda_k^{(2)}+\ihalf}{
        \lambda_j^{(3)}-\lambda_k^{(2)}-\ihalf}
  = \prod_{k =1, k\ne j}^{N_3}
  \frac{\lambda_j^{(3)}-\lambda_k^{(3)}+ i}{
        \lambda_j^{(3)}-\lambda_k^{(3)}- i}\,
  \frac{\lambda_j^{(3)}-\lambda_k^{(3)}-\ihalf}{
        \lambda_j^{(3)}-\lambda_k^{(3)}+\ihalf}\,,\quad j=1\ldots N_3\,,
  \end{aligned}
\end{equation}
where the corresponding state of the superspin chain has energy
\begin{equation}
\label{ene-o34}
  E = -L + \sum_{k=1}^{N_1} \frac{1}{\left(\lambda_k^{(1)}\right)^2+\frac14}\,.
\end{equation}

The ground state and low lying excitations of the model have root densities
$N_i/L\to 1$ in the thermodynamic limit.  We label the charge sectors of the
$OSP(3|4)$ model by quantum numbers $(n_1,n_2,n_3)= (N_1-N_2+1, N_2-N_3+1,
L-N_1-2)$.

\begin{figure}[t]
  \includegraphics[width=0.7\textwidth]{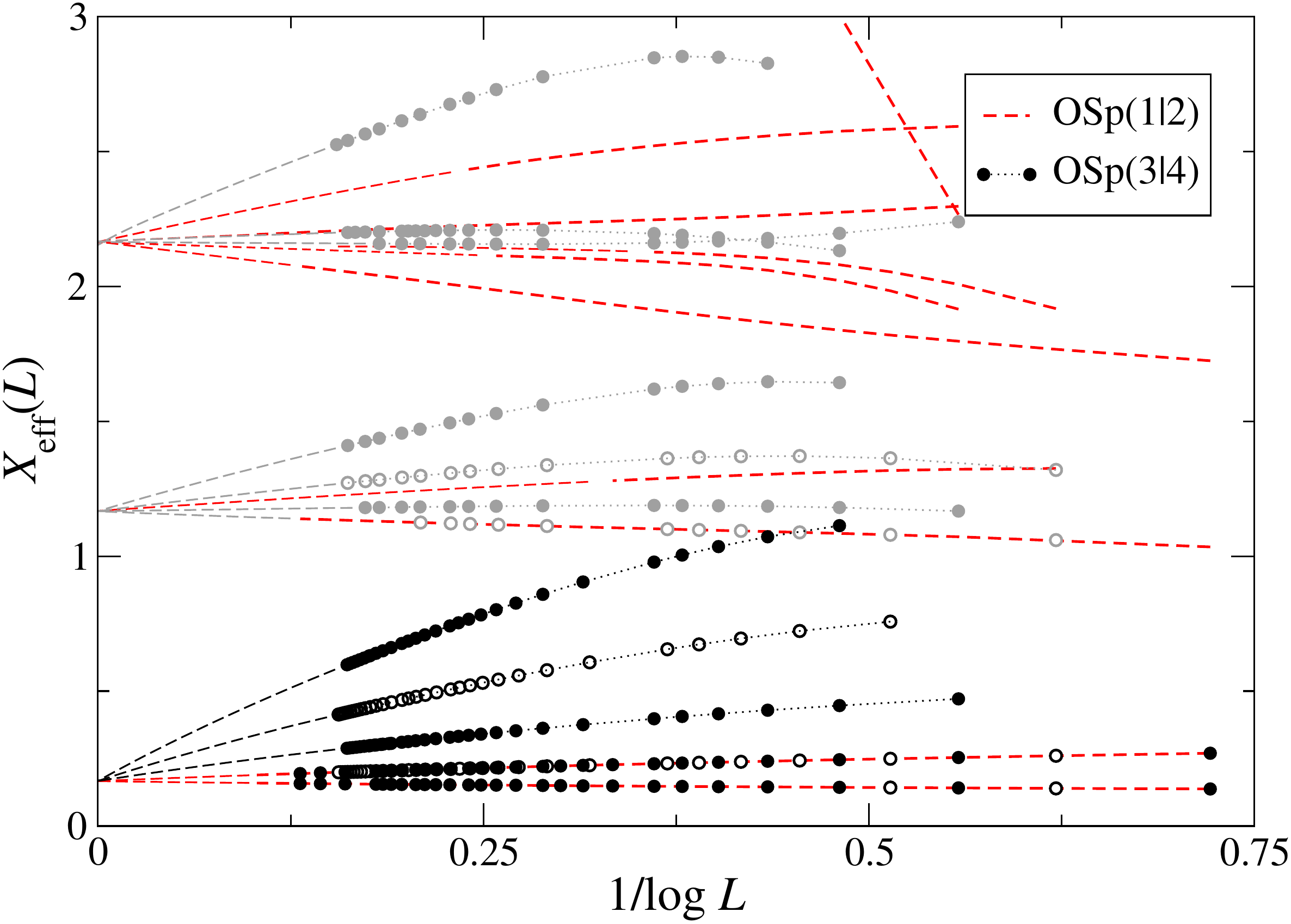}
  \caption{Finite size spectrum of the $OSp(3|4)$ (symbols) superspin chain.
    Levels already present in the $OSp(1|2)$ model due to the inclusion
    (\ref{eq:specincl}) are marked by red dashed lines.  Displayed are the
    effective scaling dimensions $X_{\mathrm{eff}}(L)$ vs.\ $1/\log L$.  Black
    symbols denote levels from the lowest tower (\ref{o34_subl}) of scaling
    dimensions in the $OSp(3|4)$ sectors $(1,1,k)$, $k=0,1,\ldots,4$.  Filled
    (open) symbols are data chains of even (odd) length.  Also shown are
    extrapolations based on a rational dependence on $1/\log
    L$.\label{fig:fso34}}
\end{figure}

The lowest energy states appear in the sectors $(n_1,n_2,n_3)=(1,1,k)$ where
$k=0,1,2,\ldots \sim L \bmod 2$.  Their Bethe roots are arranged in
$(L-k-2)/2$ complex conjugate pairs
\begin{equation*}
  \lambda_{j,\pm}^{(1)} \simeq \lambda_j^{(1)} \pm \frac{5i}{4}\,,\quad
  \lambda_{j,\pm}^{(2)} \simeq \lambda_j^{(2)} \pm \frac{3i}{4}\,,\quad
  \lambda_{j,\pm}^{(3)} \simeq \lambda_j^{(3)} \pm \frac{i}{4}\,,
\end{equation*}
and real centers $\lambda_j^{(a)}$.  In the thermodynamic limit these states
are degenerate, see Figure~\ref{fig:fso34}.
%----------------------------------------------------------------------
%----------------------------------------------------------------------
The levels with $k=0,1$ are the lowest and their energies coincide with those
of the triplet ground state and the lowest singlet excitation in the spectrum
of the $OSp(1|2)$ chain.  
Analyzing the subleading corrections to scaling for this class of levels we
conjecture ($k=0,1,2,\ldots$)
\begin{equation}
\label{o34_subl}
  X^{(3|4)}((1,1,k);L) \simeq \frac{\beta^{(3|4)}(k)}{\log L}\,,
    \quad
    \beta^{(3|4)}(k) = \frac{2k^2+2k-1}{12}\,.
\end{equation}

A second tower of primaries with spin $s=1$ levels extrapolating to
$X^{(3|4)}=1$ is found in the $OSp(3|4)$ sectors $(n_1,n_2,n_3)= (2,1,k)$ for
$k=0,1,2,\ldots \sim L-1 \bmod 2$, see Figure~\ref{fig:fso34}.  In the
corresponding Bethe root configurations one of the $N_1=L-2-k$ roots on the
first level is real.  The lowest of these excitations, $k=0$, is also present
as a triplet in the spectrum of the $OSp(1|2)$.  In addition there are
descendents of the $(1,1,k)$ primaries with scaling dimension $X=1$.

The next excitations, both of the $OSp(1|2)$ and the $OSp(3|4)$ chain, for
which we have determined the Bethe root configurations correspond to fields
with scaling dimension $X=2$.  Their conformal spin is $s=0$ or $2$.

%----------------------------------------------------------------------
\subsection{$n-2m=0$}
We study the spectrum of the $OSp(2|2)$ % \simeq sl(2|1)$
superspin chain using the Bethe equations in the grading $fbbf$:
\begin{equation}
  \label{bae-o22}
  \begin{aligned}
    &\left(\frac{\lambda_j^{(1)}+\ihalf}{\lambda_j^{(1)}-\ihalf}\right)^L
    = \prod_{k =1}^{N_-}
    \frac{\lambda_j^{(1)}-\lambda_k^{(2)}+i}{
      \lambda_j^{(1)}-\lambda_k^{(2)}-i}\,
    \quad j=1\ldots N_+\,,\\
    &\left(\frac{\lambda_j^{(2)}+\ihalf}{\lambda_j^{(2)}-\ihalf}\right)^L
    = \prod_{k =1}^{N_+}
    \frac{\lambda_j^{(2)}-\lambda_k^{(1)}+i}{
      \lambda_j^{(2)}-\lambda_k^{(1)}-i}\,
    \quad j=1\ldots N_-\,.
  \end{aligned}
\end{equation}
Solutions to these equations parametrize states with energy
\begin{equation}
  \label{ene-o22}
  E = L - \sum_{k=1}^{N_+} \frac{1}{\left(\lambda_k^{(1)}\right)^2+\frac14}
  - \sum_{k=1}^{N_-} \frac{1}{\left(\lambda_k^{(2)}\right)^2+\frac14}\,.
\end{equation}

The ground state and low energy excitations of the $OSp(2|2)$
superspin chain are described by real roots of (\ref{bae-o22}) with
densities $N_k/L\to 1/2$ in the thermodynamic limit, see
\cite{GaMa07}.  Similarly as for the $OSp(2|4)$ chain above we use
this fact to analytically compute the scaling dimensions of primary
fields from the finite size spectrum.  Introducing quantum numbers
%$n_\pm=(L/2)-N_\pm$
$n_1=L-N_+-N_-$ and $n_2=N_+-N_-$
for the $U(1)$ charges and regularizing the singularity of the Bethe ansatz
kernel we find that the scaling dimensions of primaries are
\begin{equation}
  \label{xdims-o22}
  \begin{aligned}
    X^{(2|2)}_{\mathrm{eff}} (n_k,m_k;L) \to 
    % \frac14\left((n_++n_-)^2 + {\epsilon}\,(n_+-n_-)^2 \right) 
    \frac14\left(n_1^2 + {\epsilon}\,n_2^2 \right) 
    % + \frac14\left( (m_++m_-)^2 + \frac{1}{\epsilon}(m_+-m_-)^2\right) %\right)
    + \frac14\left( m_1^2 + \frac{1}{\epsilon}m_2^2\right) %\right)
    - \frac16\,.
  \end{aligned}
\end{equation}
Their conformal spin is $s=n_1 m_1 + n_2 m_2$.  Here, the charges
$n_{1/2}$ are integers, the corresponding vorticities take values
according to the selection rules
\begin{equation}
  m_1 \sim L-n_1 \bmod 2\,, \quad m_2\in \mathbb{Z}\,.
\end{equation}
For levels from the low energy spectrum in the thermodynamic limit (where the
regularization constant $\epsilon\to 0^+$) the vorticity $m_2$ is constrained
to be $0$.

\begin{figure}[t]
  \includegraphics[width=0.7\textwidth]{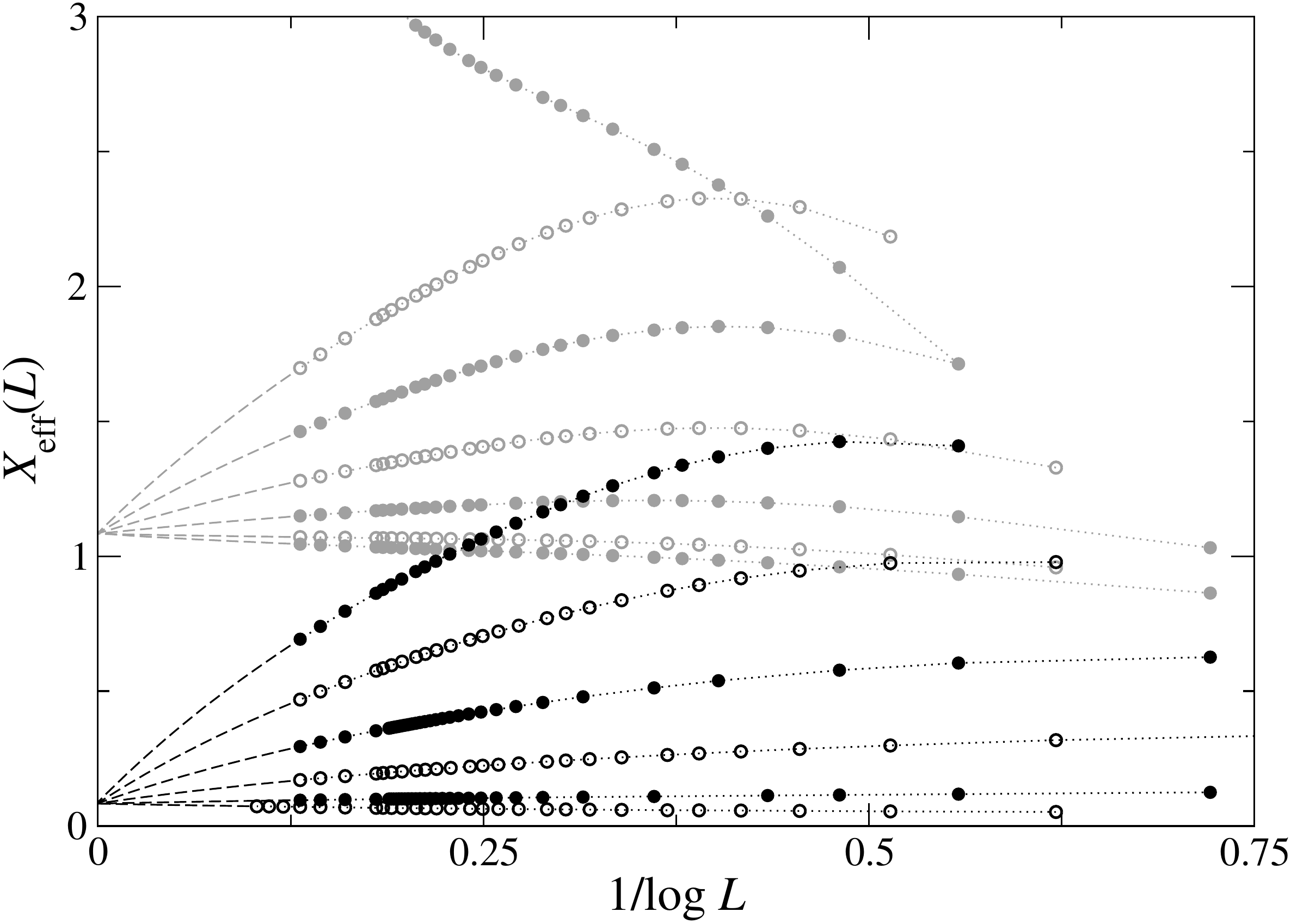}
  \caption{
    % Finite size spectrum of the $OSp(2|2)$ superspin
    % chain.  Displayed are the effective scaling dimensions
    % $X_{\mathrm{eff}}(L)$ vs.\ $1/\log L$.  Symbols are data from the solution
    % of the Bethe equations, dashed lines are extrapolations based on a
    % rational dependence on $1/\log L$.
    As Fig.~\ref{fig:fso24} but for the $OSp(2|2)$ superspin chain.  Black
    symbols denote the levels from the lowest tower (\ref{o22_subl}) of
    scaling dimensions in the sectors $(1,k)$, $k=0,1,\ldots,4$.
    \label{fig:fso22}}
\end{figure}

The lowest energy states appear in the sectors $n_1=1$, $m_1=0$ such
that
\begin{equation}
  X_{\mathrm{eff}}^{(2|2)}\left(n_1=1,m_1=0;L\right)
  \to \frac14 - \frac16 = -\frac{c_{\mathrm{eff}}}{12}\,
\end{equation}
with the effective central charge of the $OSp(2|2)$ superspin chain,
$c_{\mathrm{eff}}=-1$.  Scaling dimension and spin of the
corresponding primary are $X=0$, $s=0$.  The degeneracy of the scaling
dimensions for levels with different $n_2$ is lifted for finite system
sizes, see Figure~\ref{fig:fso22}.
Analyzing the subleading corrections to scaling of the
$(n_1,n_2)=(1,k)$ states, $k=0,1,2,\ldots \sim L-1 \bmod 2$, we find a
tower of levels:
\begin{equation}
  \label{o22_subl}
  X^{(2|2)}(\mathbf{n}=(1,k);L) \simeq \frac{\beta_k(2|2)}{\log L}\,,
  \quad
  \beta_k(2|2) = \frac{2k^2-1}{8}\,.
\end{equation}

A tower of spin $s=1$ excitations extrapolating to $X=1$ is found in
the sectors $(n_1,n_2)=(2,k)$ with $k=0,1,2,\ldots \sim L \bmod 2$.  A
state with spin $s=2$ in the sector $(2,4)$ disappears from the low
energy spectrum as the system size is increased since the restriction $m_2=0$
is violated.

%----------------------------------------------------------------------
\subsection{$n-2m=+1$}
The finite size spectrum of the $OSp(3|2)$ superspin chain has been
studied extensively using its solution by means of the algebraic Bethe
ansatz in Ref.~\cite{FrMa15}.  The ground state displays no finite
size corrections from it has been concluded that $c_{\mathrm{eff}}=0$.
The excitations considered in that work can be grouped into towers
extrapolating to integer scaling dimensions $X=0,1,2,\ldots$ or
disappear from the low energy spectrum in the thermodynamic limit.
The degeneracies in the spectrum of scaling dimensions is lifted for
finite system sizes: with the exception of the ground state the finite
size gaps show strong logarithmic corrections to scaling.  For the
levels in the $X=0$ tower these corrections have been found to scale as
\begin{equation}
  \label{o32_subl}
  X^{(3|2)}_{\mathrm{eff}}(k;L) \simeq 0 + \frac{\beta_k(3|2)}{\log L}\,,
  \quad
  \beta_k(3|2) = \frac{k(k-1)}{2}
\end{equation}

%----------------------------------------------------------------------
\subsection{$n-2m=+3$}
As mentioned above the continuum limit of the $OSp(n|2m)$ superspin chain for
$n-2m>2$ is expected to be different from that for the cases discussed so far.
Here first insights into the finite size spectrum can be obtained from the
spectral inclusion $\mathrm{spec}[O(3)] \subset \mathrm{spec}[OSp(5|2)]
\subset \ldots$, see (\ref{eq:specincl}).  The integrable $O(3)$ spin chain
(or the spin $S=1$ Takhtajan-Babujian model \cite{Takh82,Babu82}) is known to
be a lattice realization of the $SU(2)$ Wess-Zumino-Witten-Novikov (WZNW)
model at level $k=2$ with central charge $c_{\mathrm{eff}} = 3/2$ and spectrum
of conformal weights $h\in\{j(j+1)/4: j=0,1/2,1\}$.
Its primaries can be written as composite operators built from an Gaussian
representing the Kac-Moody algebra with topological charge $k=2$ and an Ising
field \cite{AlMa89}.  The lowest scaling dimensions appearing in the lattice
model of length $L$ are
\begin{equation}
  X^{O(3)}  \in \begin{cases}
    \{\frac38,1,\ldots\} & \mathrm{for~}L\mathrm{~even}\\
    \{\frac38,\frac12,1,\ldots\} & \mathrm{for~}L\mathrm{~odd}
  \end{cases}
\end{equation}
and higher descendents thereof.  We note that the level with $X^{O(3)}=1/2$
has conformal spin $s=1/2$ and is therefore not realized in the spectrum of
the even $L$ chain.
% \begin{align}
%   X^{O(3)} &= \frac14 n^2 + \frac14 m^2 + X_{\mathrm{I}}(n,m)\,,\\
%   X_{\mathrm{I}}(n,m) &\in \begin{cases}
%     0,1 & n,m\sim L \bmod 2 \\
%     \frac12 & n,m\sim L-1\bmod 2\\
%     \frac18 & n+m~\mathrm{odd}
%   \end{cases}\,.
% \end{align}
The corrections to scaling due to the marginally irrelevant perturbation of
the conformal fixed point present in the lattice Hamiltonian have been
computed in perturbation theory \cite{AGSZ89}.  For the ground state this
leads to logarithmic corrections to the central charge
\begin{equation}
  \label{o3-sublgs}
  c^{O(3)}_{\mathrm{eff}}(L) \simeq \frac32 + \frac{3}{2(\log L)^3}\,,
\end{equation}
while the finite size gap of the lowest triplet and singlet excitations are
\begin{equation}
  \label{o3-subl}
  \begin{aligned}
    X^{O(3)}_{S=1}(L) &\simeq \frac38 - \frac14\,\frac{1}{\log L}\,,\\
    X^{O(3)}_{S=0}(L) &\simeq \frac38 + \frac34\,\frac{1}{\log L}\,.
  \end{aligned}
\end{equation}
(Note that the singlet is realized in the spectrum of the $O(3)$ spin chain
with an odd number of sites only.)

The additional levels in the spectrum of the $OSp(5|2)$ superspin chain can be
studied based on its solutions by Bethe ansatz.  For the grading $fbbbbbf$ the
Bethe equations read
\begin{equation}
  \label{bae-o52}
\begin{aligned}
  &\left(\frac{\lambda_j^{(1)}+\ihalf}{\lambda_j^{(1)}-\ihalf}\right)^L
  = \prod_{k =1}^{N_2}
  \frac{\lambda_j^{(1)}-\lambda_k^{(2)}+\ihalf}{
        \lambda_j^{(1)}-\lambda_k^{(2)}-\ihalf}\,,
  &\quad j=1\ldots N_1\,,
  \\
  & \prod_{k=1,k\ne j}^{N_2}
  \frac{\lambda_j^{(2)}-\lambda_k^{(2)}+ i}{
        \lambda_j^{(2)}-\lambda_k^{(2)}- i}
  = \prod_{k =1}^{N_1}
  \frac{\lambda_j^{(2)}-\lambda_k^{(1)}+ \ihalf}{
        \lambda_j^{(2)}-\lambda_k^{(1)}- \ihalf}\,
    \prod_{k =1}^{N_3}
  \frac{\lambda_j^{(2)}-\lambda_k^{(3)}+ \ihalf}{
        \lambda_j^{(2)}-\lambda_k^{(3)}- \ihalf}\,,
     &\quad j=1\ldots N_2\,,
  \\
  & \prod_{k=1,k \ne j}^{N_3}
  \frac{\lambda_j^{(3)}-\lambda_k^{(3)}+\ihalf}{
        \lambda_j^{(3)}-\lambda_k^{(3)}-\ihalf}
  = \prod_{k =1}^{N_2}
  \frac{\lambda_j^{(3)}-\lambda_k^{(2)}+ \ihalf}{
        \lambda_j^{(3)}-\lambda_k^{(2)}- \ihalf}\,,
      & \quad j=1\ldots N_3\,.
  \end{aligned}
\end{equation}
Solutions to these equations parameterize $OSp(5|2)$ highest weight
states with energy
\begin{equation}
  E = -L + \sum_{j=1}^{N_1} \frac{1}{(\lambda_j^{(1)})^2+\frac14}\,.
\end{equation}

\begin{figure}[t]
  \includegraphics[width=0.7\textwidth]{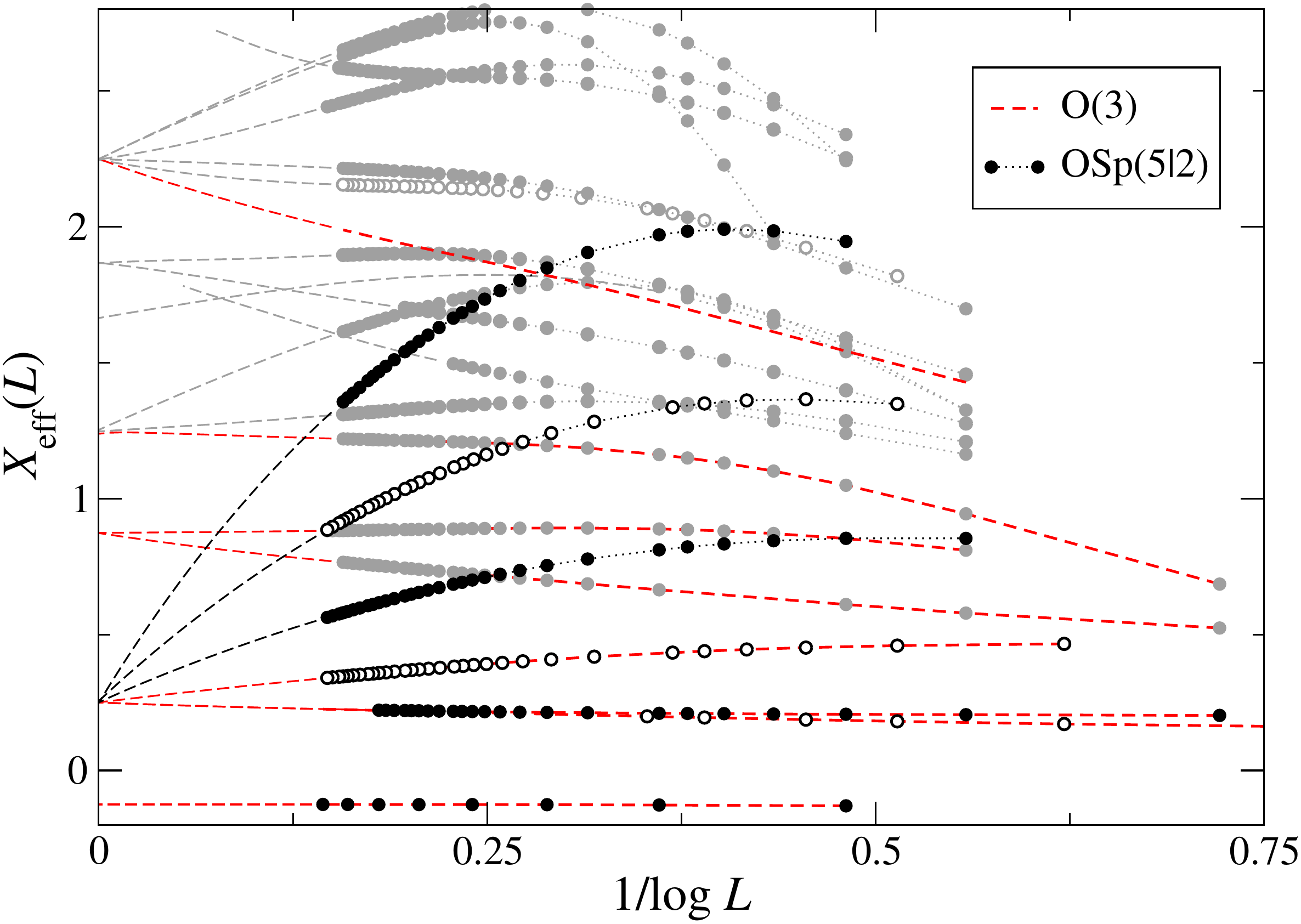}
  \caption{As Fig.~\ref{fig:fso34} but for the $OSp(5|2)$ superspin chain
    (symbols) and the $O(3)$ spin chain (red dashed lines).  Black
    symbols denote the levels from the tower (\ref{o52_subl}) of scaling
    dimensions in the $OSp(5|2)$ sectors $(k,0,0)$,
    $k=1,2,\ldots,6$.\label{fig:fso52}}
\end{figure}

Unlike in the $OSp(n|2m)$ models with $n-2m<2$ discussed above the ground
state of the superspin chain remains a unique singlet indicating the absence
of a symmetry breaking transition into a low temperature phase of the loop
models in this regime \cite{JaRS03}.  
Labeling the charge sectors of the $OSp(5|2)$ chain with quantum numbers
$(n_1,n_2,n_3) = (L-N_1,N_1-N_2,N_2-N_3)$ the lowest excitations above the
ground state of the $O(3)$ chain are found in the sector with $(k,0,0)$,
$k=1,2,3,\ldots \sim L \bmod 2$.  The corresponding Bethe root configuration
consists of $(L-k)/2$ pairs of complex conjugate roots on each level with
complex parts\footnote{For the level with $k=1$ the pairs with real part
  closest to the origin are strongly deformed.}
\begin{align*}
  \lambda_{j,\pm}^{(1)} \simeq \lambda_j^{(1)} \pm \frac{3i}{4}\,,\quad
  \lambda_{j,\pm}^{(2)} \simeq \lambda_j^{(2)} \pm \frac{i}{4}\,,\quad
  \lambda_{j,\pm}^{(3)} \simeq \lambda_j^{(3)} \pm \frac{i}{4}\,,
\end{align*}
and real centers $\lambda_j^{(a)}$, $a=1,2,3$.  The states with $k=1$ and $2$
appear as the triplet excitations in the spectrum of the $O(3)$ model for
chains of odd and even length, respectively, and the state with $k=3$ has the
energy of the $O(3)$ singlet.  The energy levels for $k>3$ are not in the
$O(3)$ part of the spectrum.  In the thermodynamic limit, $L\to\infty$, they
degenerate, see Figure~\ref{fig:fso52}.
For large but finite systems we find that this degeneracy is lifted as
\begin{equation}
  \label{o52_subl}
  X^{(5|2)}_{\mathrm{eff}}((k,0,0);L) \simeq \frac38 +
  \frac{\beta_k(5|2)}{\log L}\,, 
  \quad
  \beta_k(5|2) = \frac{2k^2-6k+3}{4}\,
\end{equation}
matching the known behaviour (\ref{o3-subl}) for the $O(3)$ levels $k=1,2,3$.

Higher energy excitations for even length superspin chains have been found
extrapolating to scaling dimensions $X=1$, $3/8+1$, $2$, and $3/8+2$, see
Figure~\ref{fig:fso52}.
The first of these is in the sector with $(n_1,n_2,n_3)=(1,1,0)$ and its
energy is that of the zero-spin field with scaling dimensions $X=1$ in the
$O(3)$ model.  Its root configuration differs from the one for the lowest
tower by one root $\lambda^{(1)}=0$ on the first level.  The energy is that of
the zero-spin field with scaling dimensions $X=1$ in the $O(3)$ model.

We have investigated the corrections to finite size scaling of mostly spin
zero levels in the $OSp(5|2)$ chain of even length up to some energy cutoff
which including the first states extrapolating to $X=3/8 + 2$.  Among these we
find no evidence for the existence of towers of dimensions exept those
starting at the descendents of the field with $X=3/8$.  This resembles the
presence of both a continuous and a discrete part in the conformal spectrum of
the $sl(2|1)$ superspin chain with alternating quark and antiquark
representations and its deformation \cite{EsFS05,SaSc07,FrMa11,FrHo17}.

\section{Discussion}
In this paper we have studied the fine structure appearing in the finite size
spectrum of the $OSp(n|2m)$ superspin chains.  We find that the ground states
of these models have a finite degeneracy for $n-2m<2$.  For large finite
system size $L$ there exists a tower of scaling dimensions extrapolating to
that of the identity operator, $X=0$, and forming a continuum in the
thermodynamic limit, $L\to\infty$.

For $n-2m=3$ (and most likely for all $n-2m>2$) the ground state of the
superspin chain is a unique singlet, with the same energy as the ground state
of the $O(3)$ spin chain.  The low energy effective description of the $O(3)$
model is known to be the $SU(2)_2$ WZNW model, its lowest two excitations with
scaling dimension $X=3/8$ show strong subleading corrections to scaling
(\ref{o3-subl}) due to the presence of a marginally irrelevant perturbation in
the lattice model.
These levels are also present in the spectrum of the $OSp(5|2)$ chain, see
Eq.~(\ref{eq:specincl}).  In addition, however, we have found continua of
scaling dimensions to emerge starting at $X=3/8$ and its descendents.

These observations are reminiscent of the appearence of a continuous component
in the spectrum of scaling dimensions in staggered (super) spin chains
\cite{EsFS05,IkJS08,FrMa11,FrMa12}.  There are some differences to our present
findings though: the fine structure in the finite size spectrum of the
staggered models has been argued to be a consequence of a non-compact degree
of freedom in the low energy theory and the subleading gaps vanish
quadratically with the inverse of $\log L$.  This has to be contrasted to the
linear dependence predicted from conformal field theory for the WZNW models
with a marginal perturbation and observed in the towers of excitations of the
$OSp(n|2m)$ models.  Similarly, the corrections to scaling of the ground state
of the staggered models due to a marginal perturbation by a \emph{continuum}
of excitations differ from (\ref{o3-sublgs}) \cite{FrSe14}.

From the analysis of our data we have formulated conjectures for the
amplitudes of the subleading (logarithmic) corrections to the lowest tower of
excitations.  For the $OSp(n|2m)$ models considered above these amplitudes are
found to be quadratic functions of the single quantum number labelling the
different levels in the lowest tower.  This suggests that they should be
connected to the quadratic Casimir of the algebra underlying the superspin
chain.
This is similar to the case of the $O(3)$ model where (\ref{o3-subl}) is
derived by studying the effect of a marginal interaction of left and right
Kac-Moody currents $\mathbf{J}_L \cdot \mathbf{J}_R$.  In a multiplet with
given left and right spin quantum numbers $\mathbf{S}_L$ and $\mathbf{S}_R$
this results in \cite{AGSZ89}
\begin{equation}
  X^{O(3)}_{\mathbf{S}_L+\mathbf{S}_R} \simeq X_{\mathrm{WZW}}  - \frac{\mathbf{S}_L\cdot
      \mathbf{S}_R}{\log L}\,.
\end{equation}

For the orthosymplectic algebras $OSp(n|2m)$ the eigenvalues of the Casimir
operator on a highest weight vector $(\nu|\mu) = (\nu_1,\ldots, \nu_{[n/2]}|
\mu_1,\ldots,\mu_m)$ are \cite{JaGr79}.
\begin{equation}
  \label{osp-casi}
  C_2^{(m|2n)}(\nu|\mu) = 2 \left\{ \sum_{a=1}^{[n/2]} \nu_a ( \nu_a + n-2m-2a)
    -\sum_{\alpha=1}^{m} \mu_a ( \mu_a + 2m + \ell_n
    -2\alpha)\right\} \,,
\end{equation}
where $\ell_n = 2$ ($1$) for $n$ even (odd).  The first term in this
expression, $2\nu_1(\nu_1+n-2m-2)$, is present in the Casimir for each of the
algebras with given $n-2m$ for $n>1$.  We note that, the amplitudes of the
subleading logarithmic corrections measured relative to the first level with
positive scaling dimension (i.e.\ after subtracting the smallest non-negative
amplitude $\beta^{(n|2m)}$) appearing in the superspin chains with $n-2m<2$
for even $L$ the amplitudes can be directly related to the corresponding
Casimir eigenvalue:
\begin{align*}
  OSp(2|4): \quad \beta^{(2|4)}(k) - \frac38 &= \frac{(k+2)(k-2)}{8} =
  \frac{C_2^{(2|4)}(k+2|0,0) }{16}\,,\\
  OSp(3|4): \quad \beta^{(3|4)}(k) - \frac14 &= \frac{(k+2)(k-1)}{6} =
  \frac{C_2^{(3|4)}(k+2|0,0) }{12}\,,\\
  OSp(2|2): \quad \beta^{(2|2)}(k) - \frac18 &= \frac{(k+1)(k-1)}{4} =
  \frac{C_2^{(2|2)}(k+1|0) }{8}\,,\\
  OSp(3|2): \quad \beta^{(3|2)}(k) - 0 &= \frac{k(k-1)}{2} =
  \frac{C_2^{(3|2)}(k|0) }{4}\,.
\end{align*}
This can be compared to the case of $OSp(5|2)$ where the finite part of the
scaling dimension for the lowest tower is already positive.  Therefore,
measuring $\beta^{(5|2)}$ relative to the smallest one we get
\begin{align*}
  OSp(5|2): \quad \beta^{(5|2)}(k) + \frac14 = \frac{(k-1)(k-2)}{2} = 
  \frac{C_2^{(5|2)}(k-2|0,0) }{4}\,.
\end{align*}

Being corrections to the scaling dimensions of primary field these amplitudes
are expected to determine logarithmic corrections to correlation functions.
Again, the scaling dimensions should be measured starting from the smallest
non-negative one for the given model.  For the $OSp(n|2m)$ models with
$n-2m<2$ this predicts two-point functions of these fields to be
\begin{equation}
  \label{corr-watermelon}
  G_k^{(n|2m)}(r) \sim (\log r)^{-\alpha_k}\,,\qquad
  \alpha_k = \frac{k(k+n-2m-2)}{2-n+2m}\,
\end{equation}
For the correlation functions to decay at large distances $k$ has to be
restricted to $k>2-n+2m$.  Eq.~(\ref{corr-watermelon}) agrees with the
exponents for $k$-leg watermelon correlators proposed for the Goldstone phase
of intersecting loop models with fugacity $n-2m<2$ and studied numerically
using Monte Carlo simulations for $n-2m=1$ \cite{Polyakov75,NSSO13}.

Eq.~(\ref{corr-watermelon}) leads us to propose the existence of a family of
fields in the $OSp(5|2)$ model whose two-point correlation functions feature a
multiplicative logarithmic correction to the algebraic behaviour expected from
conformal field theory, i.e.\
\begin{equation}
  G_k^{(5|2)}(r) \sim \frac{1}{r^{3/4}}\, (\log r)^{-k^2-k+1/2}\, \quad k\ge 0\,.
\end{equation}
We note that $G_{k=0}^{(5|2)}(r)$ is the spin-spin correlation function in the
$O(3)$ model \cite{AGSZ89}.

%%%%%%%%%%%%%%%%%%%%%%%%%%%%%%%%%%%%%%%%%%%%%%%%%%%%%%%%%%%%%%%%%%%%%%
\begin{acknowledgments}
  This work has been carried out within the research unit \emph{Correlations
    in Integrable Quantum Many-Body Systems} (FOR2316).  Financial support by
  the Deutsche Forschungsgemeinschaft through grant No.\ Fr~737/9-1 is
  gratefully acknowledged.  M.J.M.\ thanks the Brazilian Research Council CNPq
  for funding through grant no.\ 2016/401694.
\end{acknowledgments}

%\bibliography{base,bound,books,frahm}

\begin{thebibliography}{37}%
\makeatletter
\providecommand \@ifxundefined [1]{%
 \@ifx{#1\undefined}
}%
\providecommand \@ifnum [1]{%
 \ifnum #1\expandafter \@firstoftwo
 \else \expandafter \@secondoftwo
 \fi
}%
\providecommand \@ifx [1]{%
 \ifx #1\expandafter \@firstoftwo
 \else \expandafter \@secondoftwo
 \fi
}%
\providecommand \natexlab [1]{#1}%
\providecommand \enquote  [1]{``#1''}%
\providecommand \bibnamefont  [1]{#1}%
\providecommand \bibfnamefont [1]{#1}%
\providecommand \citenamefont [1]{#1}%
\providecommand \href@noop [0]{\@secondoftwo}%
\providecommand \href [0]{\begingroup \@sanitize@url \@href}%
\providecommand \@href[1]{\@@startlink{#1}\@@href}%
\providecommand \@@href[1]{\endgroup#1\@@endlink}%
\providecommand \@sanitize@url [0]{\catcode `\\12\catcode `\$12\catcode
  `\&12\catcode `\#12\catcode `\^12\catcode `\_12\catcode `\%12\relax}%
\providecommand \@@startlink[1]{}%
\providecommand \@@endlink[0]{}%
\providecommand \url  [0]{\begingroup\@sanitize@url \@url }%
\providecommand \@url [1]{\endgroup\@href {#1}{\urlprefix }}%
\providecommand \urlprefix  [0]{URL }%
\providecommand \Eprint [0]{\href }%
\providecommand \doibase [0]{http://dx.doi.org/}%
\providecommand \selectlanguage [0]{\@gobble}%
\providecommand \bibinfo  [0]{\@secondoftwo}%
\providecommand \bibfield  [0]{\@secondoftwo}%
\providecommand \translation [1]{[#1]}%
\providecommand \BibitemOpen [0]{}%
\providecommand \bibitemStop [0]{}%
\providecommand \bibitemNoStop [0]{.\EOS\space}%
\providecommand \EOS [0]{\spacefactor3000\relax}%
\providecommand \BibitemShut  [1]{\csname bibitem#1\endcsname}%
\let\auto@bib@innerbib\@empty
%</preamble>
\bibitem [{\citenamefont {Cardy}(1987)}]{Cardy87}%
  \BibitemOpen
  \bibfield  {author} {\bibinfo {author} {\bibfnamefont {J.~L.}\ \bibnamefont
  {Cardy}},\ }in\ \href@noop {} {\emph {\bibinfo {booktitle} {Phase Transitions
  and Critical Phenomena}}},\ Vol.~\bibinfo {volume} {11},\ \bibinfo {editor}
  {edited by\ \bibinfo {editor} {\bibfnamefont {C.}~\bibnamefont {Domb}}\ and\
  \bibinfo {editor} {\bibfnamefont {J.~L.}\ \bibnamefont {Lebowitz}}}\
  (\bibinfo  {publisher} {Academic Press},\ \bibinfo {address} {London},\
  \bibinfo {year} {1987})\ pp.\ \bibinfo {pages} {55--126}\BibitemShut
  {NoStop}%
\bibitem [{\citenamefont {Saleur}(2000)}]{Saleur00}%
  \BibitemOpen
  \bibfield  {author} {\bibinfo {author} {\bibfnamefont {H.}~\bibnamefont
  {Saleur}},\ }\href@noop {} {\bibfield  {journal} {\bibinfo  {journal} {Nucl.
  Phys. B}\ }\textbf {\bibinfo {volume} {578}},\ \bibinfo {pages} {552}
  (\bibinfo {year} {2000})},\ \Eprint {http://arxiv.org/abs/solv-int/9905007}
  {solv-int/9905007} \BibitemShut {NoStop}%
\bibitem [{\citenamefont {Martins}\ \emph {et~al.}(1998)\citenamefont
  {Martins}, \citenamefont {Nienhuis},\ and\ \citenamefont {Rietman}}]{MaNR98}%
  \BibitemOpen
  \bibfield  {author} {\bibinfo {author} {\bibfnamefont {M.~J.}\ \bibnamefont
  {Martins}}, \bibinfo {author} {\bibfnamefont {B.}~\bibnamefont {Nienhuis}}, \
  and\ \bibinfo {author} {\bibfnamefont {R.}~\bibnamefont {Rietman}},\
  }\href@noop {} {\bibfield  {journal} {\bibinfo  {journal} {Phys. Rev. Lett.}\
  }\textbf {\bibinfo {volume} {81}},\ \bibinfo {pages} {504} (\bibinfo {year}
  {1998})},\ \Eprint {http://arxiv.org/abs/cond-mat/9709051} {cond-mat/9709051}
  \BibitemShut {NoStop}%
\bibitem [{\citenamefont {Frahm}\ and\ \citenamefont {Martins}(2015)}]{FrMa15}%
  \BibitemOpen
  \bibfield  {author} {\bibinfo {author} {\bibfnamefont {H.}~\bibnamefont
  {Frahm}}\ and\ \bibinfo {author} {\bibfnamefont {M.~J.}\ \bibnamefont
  {Martins}},\ }\href@noop {} {\bibfield  {journal} {\bibinfo  {journal} {Nucl.
  Phys. B}\ }\textbf {\bibinfo {volume} {894}},\ \bibinfo {pages} {665}
  (\bibinfo {year} {2015})},\ \Eprint {http://arxiv.org/abs/1502.05305}
  {arXiv:1502.05305} \BibitemShut {NoStop}%
\bibitem [{\citenamefont {Essler}\ \emph {et~al.}(2005)\citenamefont {Essler},
  \citenamefont {Frahm},\ and\ \citenamefont {Saleur}}]{EsFS05}%
  \BibitemOpen
  \bibfield  {author} {\bibinfo {author} {\bibfnamefont {F.~H.~L.}\
  \bibnamefont {Essler}}, \bibinfo {author} {\bibfnamefont {H.}~\bibnamefont
  {Frahm}}, \ and\ \bibinfo {author} {\bibfnamefont {H.}~\bibnamefont
  {Saleur}},\ }\href@noop {} {\bibfield  {journal} {\bibinfo  {journal} {Nucl.
  Phys. B}\ }\textbf {\bibinfo {volume} {712 [FS]}},\ \bibinfo {pages} {513}
  (\bibinfo {year} {2005})},\ \Eprint {http://arxiv.org/abs/cond-mat/0501197}
  {cond-mat/0501197} \BibitemShut {NoStop}%
\bibitem [{\citenamefont {Ikhlef}\ \emph {et~al.}(2008)\citenamefont {Ikhlef},
  \citenamefont {Jacobsen},\ and\ \citenamefont {Saleur}}]{IkJS08}%
  \BibitemOpen
  \bibfield  {author} {\bibinfo {author} {\bibfnamefont {Y.}~\bibnamefont
  {Ikhlef}}, \bibinfo {author} {\bibfnamefont {J.~L.}\ \bibnamefont
  {Jacobsen}}, \ and\ \bibinfo {author} {\bibfnamefont {H.}~\bibnamefont
  {Saleur}},\ }\href@noop {} {\bibfield  {journal} {\bibinfo  {journal} {Nucl.
  Phys. B}\ }\textbf {\bibinfo {volume} {789}},\ \bibinfo {pages} {483}
  (\bibinfo {year} {2008})},\ \Eprint {http://arxiv.org/abs/cond-mat/0612037}
  {cond-mat/0612037} \BibitemShut {NoStop}%
\bibitem [{\citenamefont {Jacobsen}\ \emph {et~al.}(2003)\citenamefont
  {Jacobsen}, \citenamefont {Read},\ and\ \citenamefont {Saleur}}]{JaRS03}%
  \BibitemOpen
  \bibfield  {author} {\bibinfo {author} {\bibfnamefont {J.~L.}\ \bibnamefont
  {Jacobsen}}, \bibinfo {author} {\bibfnamefont {N.}~\bibnamefont {Read}}, \
  and\ \bibinfo {author} {\bibfnamefont {H.}~\bibnamefont {Saleur}},\
  }\href@noop {} {\bibfield  {journal} {\bibinfo  {journal} {Phys. Rev. Lett.}\
  }\textbf {\bibinfo {volume} {90}},\ \bibinfo {pages} {090601} (\bibinfo
  {year} {2003})},\ \Eprint {http://arxiv.org/abs/cond-mat/0205033}
  {cond-mat/0205033} \BibitemShut {NoStop}%
\bibitem [{\citenamefont {Polyakov}(1975)}]{Polyakov75}%
  \BibitemOpen
  \bibfield  {author} {\bibinfo {author} {\bibfnamefont {A.~M.}\ \bibnamefont
  {Polyakov}},\ }\href@noop {} {\bibfield  {journal} {\bibinfo  {journal}
  {Phys. Lett. B}\ }\textbf {\bibinfo {volume} {59}},\ \bibinfo {pages} {79}
  (\bibinfo {year} {1975})}\BibitemShut {NoStop}%
\bibitem [{\citenamefont {Nahum}\ \emph {et~al.}(2013)\citenamefont {Nahum},
  \citenamefont {Serna}, \citenamefont {Somoza},\ and\ \citenamefont
  {Ortu{\~n}o}}]{NSSO13}%
  \BibitemOpen
  \bibfield  {author} {\bibinfo {author} {\bibfnamefont {A.}~\bibnamefont
  {Nahum}}, \bibinfo {author} {\bibfnamefont {P.}~\bibnamefont {Serna}},
  \bibinfo {author} {\bibfnamefont {A.~M.}\ \bibnamefont {Somoza}}, \ and\
  \bibinfo {author} {\bibfnamefont {M.}~\bibnamefont {Ortu{\~n}o}},\
  }\href@noop {} {\bibfield  {journal} {\bibinfo  {journal} {Phys. Rev. B}\
  }\textbf {\bibinfo {volume} {87}},\ \bibinfo {pages} {184204} (\bibinfo
  {year} {2013})},\ \Eprint {http://arxiv.org/abs/1303.2342} {arXiv:1303.2342}
  \BibitemShut {NoStop}%
\bibitem [{\citenamefont {Kulish}(1986)}]{Kulish85}%
  \BibitemOpen
  \bibfield  {author} {\bibinfo {author} {\bibfnamefont {P.~P.}\ \bibnamefont
  {Kulish}},\ }\href@noop {} {\bibfield  {journal} {\bibinfo  {journal} {J.
  Sov. Math.}\ }\textbf {\bibinfo {volume} {35}},\ \bibinfo {pages} {2648}
  (\bibinfo {year} {1986})},\ \bibinfo {note} {[translated from Zap. Nauch.
  Semin. LOMI {\bf 145}, 140 (1985)]}\BibitemShut {NoStop}%
\bibitem [{\citenamefont {Martins}\ and\ \citenamefont
  {Ramos}(1997)}]{MaRa97a}%
  \BibitemOpen
  \bibfield  {author} {\bibinfo {author} {\bibfnamefont {M.~J.}\ \bibnamefont
  {Martins}}\ and\ \bibinfo {author} {\bibfnamefont {P.~B.}\ \bibnamefont
  {Ramos}},\ }\href@noop {} {\bibfield  {journal} {\bibinfo  {journal} {Nucl.
  Phys. B}\ }\textbf {\bibinfo {volume} {500}},\ \bibinfo {pages} {579}
  (\bibinfo {year} {1997})},\ \Eprint {http://arxiv.org/abs/hep-th/9703023}
  {hep-th/9703023} \BibitemShut {NoStop}%
\bibitem [{\citenamefont {Stroganov}(1979)}]{Stro79}%
  \BibitemOpen
  \bibfield  {author} {\bibinfo {author} {\bibfnamefont {{\relax Yu}.~G.}\
  \bibnamefont {Stroganov}},\ }\href@noop {} {\bibfield  {journal} {\bibinfo
  {journal} {Phys. Lett.}\ }\textbf {\bibinfo {volume} {74A}},\ \bibinfo
  {pages} {116} (\bibinfo {year} {1979})}\BibitemShut {NoStop}%
\bibitem [{\citenamefont {Baxter}(1982)}]{Baxt82a}%
  \BibitemOpen
  \bibfield  {author} {\bibinfo {author} {\bibfnamefont {R.~J.}\ \bibnamefont
  {Baxter}},\ }\href@noop {} {\bibfield  {journal} {\bibinfo  {journal} {J.
  Stat. Phys.}\ }\textbf {\bibinfo {volume} {28}},\ \bibinfo {pages} {1}
  (\bibinfo {year} {1982})}\BibitemShut {NoStop}%
\bibitem [{\citenamefont {Kl{\"u}mper}(1990)}]{Klumper90}%
  \BibitemOpen
  \bibfield  {author} {\bibinfo {author} {\bibfnamefont {A.}~\bibnamefont
  {Kl{\"u}mper}},\ }\href@noop {} {\bibfield  {journal} {\bibinfo  {journal}
  {J. Phys. A: Math. Gen.}\ }\textbf {\bibinfo {volume} {23}},\ \bibinfo
  {pages} {809} (\bibinfo {year} {1990})}\BibitemShut {NoStop}%
\bibitem [{\citenamefont {Candu}(2011)}]{Cand11}%
  \BibitemOpen
  \bibfield  {author} {\bibinfo {author} {\bibfnamefont {C.}~\bibnamefont
  {Candu}},\ }\href@noop {} {\bibfield  {journal} {\bibinfo  {journal} {JHEP}\
  }\textbf {\bibinfo {volume} {1107}},\ \bibinfo {pages} {69} (\bibinfo {year}
  {2011})},\ \Eprint {http://arxiv.org/abs/1012.0050} {arXiv:1012.0050}
  \BibitemShut {NoStop}%
\bibitem [{\citenamefont {Martins}(1990)}]{Martins90a}%
  \BibitemOpen
  \bibfield  {author} {\bibinfo {author} {\bibfnamefont {M.~J.}\ \bibnamefont
  {Martins}},\ }\href@noop {} {\bibfield  {journal} {\bibinfo  {journal} {Phys.
  Lett. A}\ }\textbf {\bibinfo {volume} {145}},\ \bibinfo {pages} {127}
  (\bibinfo {year} {1990})}\BibitemShut {NoStop}%
\bibitem [{\citenamefont {Martins}(1991)}]{Martins91}%
  \BibitemOpen
  \bibfield  {author} {\bibinfo {author} {\bibfnamefont {M.~J.}\ \bibnamefont
  {Martins}},\ }\href@noop {} {\bibfield  {journal} {\bibinfo  {journal} {J.
  Phys. A: Math. Gen.}\ }\textbf {\bibinfo {volume} {24}},\ \bibinfo {pages}
  {L159} (\bibinfo {year} {1991})}\BibitemShut {NoStop}%
\bibitem [{\citenamefont {Bl{\"o}te}\ \emph {et~al.}(1986)\citenamefont
  {Bl{\"o}te}, \citenamefont {Cardy},\ and\ \citenamefont
  {Nightingale}}]{BlCN86}%
  \BibitemOpen
  \bibfield  {author} {\bibinfo {author} {\bibfnamefont {H.~W.~J.}\
  \bibnamefont {Bl{\"o}te}}, \bibinfo {author} {\bibfnamefont {J.~L.}\
  \bibnamefont {Cardy}}, \ and\ \bibinfo {author} {\bibfnamefont {M.~P.}\
  \bibnamefont {Nightingale}},\ }\href@noop {} {\bibfield  {journal} {\bibinfo
  {journal} {Phys. Rev. Lett.}\ }\textbf {\bibinfo {volume} {56}},\ \bibinfo
  {pages} {742} (\bibinfo {year} {1986})}\BibitemShut {NoStop}%
\bibitem [{\citenamefont {Affleck}(1986)}]{Affl86}%
  \BibitemOpen
  \bibfield  {author} {\bibinfo {author} {\bibfnamefont {I.}~\bibnamefont
  {Affleck}},\ }\href@noop {} {\bibfield  {journal} {\bibinfo  {journal} {Phys.
  Rev. Lett.}\ }\textbf {\bibinfo {volume} {56}},\ \bibinfo {pages} {746}
  (\bibinfo {year} {1986})}\BibitemShut {NoStop}%
\bibitem [{\citenamefont {Cardy}(1986{\natexlab{a}})}]{Card86a}%
  \BibitemOpen
  \bibfield  {author} {\bibinfo {author} {\bibfnamefont {J.~L.}\ \bibnamefont
  {Cardy}},\ }\href@noop {} {\bibfield  {journal} {\bibinfo  {journal} {Nucl.
  Phys. B}\ }\textbf {\bibinfo {volume} {270}},\ \bibinfo {pages} {186}
  (\bibinfo {year} {1986}{\natexlab{a}})}\BibitemShut {NoStop}%
\bibitem [{\citenamefont {Cardy}(1986{\natexlab{b}})}]{Card86c}%
  \BibitemOpen
  \bibfield  {author} {\bibinfo {author} {\bibfnamefont {J.~L.}\ \bibnamefont
  {Cardy}},\ }\href@noop {} {\bibfield  {journal} {\bibinfo  {journal} {J.
  Phys. A: Math. Gen.}\ }\textbf {\bibinfo {volume} {19}},\ \bibinfo {pages}
  {L1093} (\bibinfo {year} {1986}{\natexlab{b}})}\BibitemShut {NoStop}%
\bibitem [{\citenamefont {Sogo}(1984)}]{Sogo84}%
  \BibitemOpen
  \bibfield  {author} {\bibinfo {author} {\bibfnamefont {K.}~\bibnamefont
  {Sogo}},\ }\href@noop {} {\bibfield  {journal} {\bibinfo  {journal} {Phys.
  Lett. A}\ }\textbf {\bibinfo {volume} {104}},\ \bibinfo {pages} {51}
  (\bibinfo {year} {1984})}\BibitemShut {NoStop}%
\bibitem [{\citenamefont {de~Vega}\ and\ \citenamefont
  {Woynarowich}(1985)}]{VeWo85}%
  \BibitemOpen
  \bibfield  {author} {\bibinfo {author} {\bibfnamefont {H.~J.}\ \bibnamefont
  {de~Vega}}\ and\ \bibinfo {author} {\bibfnamefont {F.}~\bibnamefont
  {Woynarowich}},\ }\href@noop {} {\bibfield  {journal} {\bibinfo  {journal}
  {Nucl. Phys. B}\ }\textbf {\bibinfo {volume} {251}},\ \bibinfo {pages} {439}
  (\bibinfo {year} {1985})}\BibitemShut {NoStop}%
\bibitem [{\citenamefont {Bogoliubov}\ \emph {et~al.}(1986)\citenamefont
  {Bogoliubov}, \citenamefont {Izergin},\ and\ \citenamefont
  {Korepin}}]{BoIK86}%
  \BibitemOpen
  \bibfield  {author} {\bibinfo {author} {\bibfnamefont {N.~M.}\ \bibnamefont
  {Bogoliubov}}, \bibinfo {author} {\bibfnamefont {A.~G.}\ \bibnamefont
  {Izergin}}, \ and\ \bibinfo {author} {\bibfnamefont {V.~E.}\ \bibnamefont
  {Korepin}},\ }\href@noop {} {\bibfield  {journal} {\bibinfo  {journal} {Nucl.
  Phys. B}\ }\textbf {\bibinfo {volume} {275 [FS17]}},\ \bibinfo {pages} {687}
  (\bibinfo {year} {1986})}\BibitemShut {NoStop}%
\bibitem [{\citenamefont {Frahm}\ and\ \citenamefont {Martins}(2011)}]{FrMa11}%
  \BibitemOpen
  \bibfield  {author} {\bibinfo {author} {\bibfnamefont {H.}~\bibnamefont
  {Frahm}}\ and\ \bibinfo {author} {\bibfnamefont {M.~J.}\ \bibnamefont
  {Martins}},\ }\href@noop {} {\bibfield  {journal} {\bibinfo  {journal} {Nucl.
  Phys. B}\ }\textbf {\bibinfo {volume} {847}},\ \bibinfo {pages} {220}
  (\bibinfo {year} {2011})},\ \Eprint {http://arxiv.org/abs/1012.1753}
  {arXiv:1012.1753} \BibitemShut {NoStop}%
\bibitem [{\citenamefont {Frahm}\ and\ \citenamefont {Martins}(2012)}]{FrMa12}%
  \BibitemOpen
  \bibfield  {author} {\bibinfo {author} {\bibfnamefont {H.}~\bibnamefont
  {Frahm}}\ and\ \bibinfo {author} {\bibfnamefont {M.~J.}\ \bibnamefont
  {Martins}},\ }\href@noop {} {\bibfield  {journal} {\bibinfo  {journal} {Nucl.
  Phys. B}\ }\textbf {\bibinfo {volume} {862}},\ \bibinfo {pages} {504}
  (\bibinfo {year} {2012})},\ \Eprint {http://arxiv.org/abs/1202.4676}
  {arXiv:1202.4676} \BibitemShut {NoStop}%
\bibitem [{\citenamefont {Martins}(1995)}]{Mart95}%
  \BibitemOpen
  \bibfield  {author} {\bibinfo {author} {\bibfnamefont {M.~J.}\ \bibnamefont
  {Martins}},\ }\href@noop {} {\bibfield  {journal} {\bibinfo  {journal} {Nucl.
  Phys. B}\ }\textbf {\bibinfo {volume} {450}},\ \bibinfo {pages} {768}
  (\bibinfo {year} {1995})},\ \Eprint {http://arxiv.org/abs/hep-th/9502133}
  {hep-th/9502133} \BibitemShut {NoStop}%
\bibitem [{\citenamefont {Tavares}\ and\ \citenamefont
  {Ribeiro}(2017)}]{TaRi17}%
  \BibitemOpen
  \bibfield  {author} {\bibinfo {author} {\bibfnamefont {T.~S.}\ \bibnamefont
  {Tavares}}\ and\ \bibinfo {author} {\bibfnamefont {G.~A.~P.}\ \bibnamefont
  {Ribeiro}},\ }\href {\doibase 10.1016/j.nuclphysb.2017.05.020} {\bibfield
  {journal} {\bibinfo  {journal} {Nucl. Phys. B}\ }\textbf {\bibinfo {volume}
  {921}},\ \bibinfo {pages} {357} (\bibinfo {year} {2017})},\ \Eprint
  {http://arxiv.org/abs/1702.08086v1} {arXiv:1702.08086v1} \BibitemShut
  {NoStop}%
\bibitem [{\citenamefont {Galleas}\ and\ \citenamefont
  {Martins}(2007)}]{GaMa07}%
  \BibitemOpen
  \bibfield  {author} {\bibinfo {author} {\bibfnamefont {W.}~\bibnamefont
  {Galleas}}\ and\ \bibinfo {author} {\bibfnamefont {M.~J.}\ \bibnamefont
  {Martins}},\ }\href@noop {} {\bibfield  {journal} {\bibinfo  {journal} {Nucl.
  Phys. B}\ }\textbf {\bibinfo {volume} {768}},\ \bibinfo {pages} {219}
  (\bibinfo {year} {2007})},\ \Eprint {http://arxiv.org/abs/hep-th/0612281}
  {hep-th/0612281} \BibitemShut {NoStop}%
\bibitem [{\citenamefont {Takhtajan}(1982)}]{Takh82}%
  \BibitemOpen
  \bibfield  {author} {\bibinfo {author} {\bibfnamefont {L.}~\bibnamefont
  {Takhtajan}},\ }\href@noop {} {\bibfield  {journal} {\bibinfo  {journal}
  {Phys. Lett. A}\ }\textbf {\bibinfo {volume} {87}},\ \bibinfo {pages} {479}
  (\bibinfo {year} {1982})}\BibitemShut {NoStop}%
\bibitem [{\citenamefont {Babujian}(1982)}]{Babu82}%
  \BibitemOpen
  \bibfield  {author} {\bibinfo {author} {\bibfnamefont {H.~M.}\ \bibnamefont
  {Babujian}},\ }\href@noop {} {\bibfield  {journal} {\bibinfo  {journal}
  {Phys. Lett. A}\ }\textbf {\bibinfo {volume} {90}},\ \bibinfo {pages} {479}
  (\bibinfo {year} {1982})}\BibitemShut {NoStop}%
\bibitem [{\citenamefont {Alcaraz}\ and\ \citenamefont
  {Martins}(1989)}]{AlMa89}%
  \BibitemOpen
  \bibfield  {author} {\bibinfo {author} {\bibfnamefont {F.~C.}\ \bibnamefont
  {Alcaraz}}\ and\ \bibinfo {author} {\bibfnamefont {M.~J.}\ \bibnamefont
  {Martins}},\ }\href@noop {} {\bibfield  {journal} {\bibinfo  {journal} {J.
  Phys. A: Math. Gen.}\ }\textbf {\bibinfo {volume} {22}},\ \bibinfo {pages}
  {1829} (\bibinfo {year} {1989})}\BibitemShut {NoStop}%
\bibitem [{\citenamefont {Affleck}\ \emph {et~al.}(1989)\citenamefont
  {Affleck}, \citenamefont {Gepner}, \citenamefont {Schulz},\ and\
  \citenamefont {Ziman}}]{AGSZ89}%
  \BibitemOpen
  \bibfield  {author} {\bibinfo {author} {\bibfnamefont {I.}~\bibnamefont
  {Affleck}}, \bibinfo {author} {\bibfnamefont {D.}~\bibnamefont {Gepner}},
  \bibinfo {author} {\bibfnamefont {H.~J.}\ \bibnamefont {Schulz}}, \ and\
  \bibinfo {author} {\bibfnamefont {T.}~\bibnamefont {Ziman}},\ }\href@noop {}
  {\bibfield  {journal} {\bibinfo  {journal} {J. Phys. A: Math. Gen.}\ }\textbf
  {\bibinfo {volume} {22}},\ \bibinfo {pages} {511} (\bibinfo {year}
  {1989})}\BibitemShut {NoStop}%
\bibitem [{\citenamefont {Saleur}\ and\ \citenamefont
  {Schomerus}(2007)}]{SaSc07}%
  \BibitemOpen
  \bibfield  {author} {\bibinfo {author} {\bibfnamefont {H.}~\bibnamefont
  {Saleur}}\ and\ \bibinfo {author} {\bibfnamefont {V.}~\bibnamefont
  {Schomerus}},\ }\href@noop {} {\bibfield  {journal} {\bibinfo  {journal}
  {Nucl. Phys. B}\ }\textbf {\bibinfo {volume} {775}},\ \bibinfo {pages} {312}
  (\bibinfo {year} {2007})},\ \Eprint {http://arxiv.org/abs/hep-th/0611147}
  {hep-th/0611147} \BibitemShut {NoStop}%
\bibitem [{\citenamefont {Frahm}\ and\ \citenamefont
  {Hobu{\ss}}(2017)}]{FrHo17}%
  \BibitemOpen
  \bibfield  {author} {\bibinfo {author} {\bibfnamefont {H.}~\bibnamefont
  {Frahm}}\ and\ \bibinfo {author} {\bibfnamefont {K.}~\bibnamefont
  {Hobu{\ss}}},\ }\href {\doibase 10.1088/1751-8121/aa77e7} {\bibfield
  {journal} {\bibinfo  {journal} {J. Phys. A: Math. Theor.}\ }\textbf {\bibinfo
  {volume} {50}},\ \bibinfo {pages} {294002} (\bibinfo {year} {2017})},\
  \Eprint {http://arxiv.org/abs/1703.08054} {1703.08054} \BibitemShut {NoStop}%
\bibitem [{\citenamefont {Frahm}\ and\ \citenamefont {Seel}(2014)}]{FrSe14}%
  \BibitemOpen
  \bibfield  {author} {\bibinfo {author} {\bibfnamefont {H.}~\bibnamefont
  {Frahm}}\ and\ \bibinfo {author} {\bibfnamefont {A.}~\bibnamefont {Seel}},\
  }\href@noop {} {\bibfield  {journal} {\bibinfo  {journal} {Nucl. Phys. B}\
  }\textbf {\bibinfo {volume} {879}},\ \bibinfo {pages} {382} (\bibinfo {year}
  {2014})},\ \Eprint {http://arxiv.org/abs/1311.6911} {arXiv:1311.6911}
  \BibitemShut {NoStop}%
\bibitem [{\citenamefont {Jarvis}\ and\ \citenamefont {Green}(1979)}]{JaGr79}%
  \BibitemOpen
  \bibfield  {author} {\bibinfo {author} {\bibfnamefont {P.~D.}\ \bibnamefont
  {Jarvis}}\ and\ \bibinfo {author} {\bibfnamefont {H.~S.}\ \bibnamefont
  {Green}},\ }\href@noop {} {\bibfield  {journal} {\bibinfo  {journal} {J.
  Math. Phys.}\ }\textbf {\bibinfo {volume} {20}},\ \bibinfo {pages} {2115}
  (\bibinfo {year} {1979})}\BibitemShut {NoStop}%
\end{thebibliography}
 %merlin.mbs apsrev4-1.bst 2010-07-25 4.21a (PWD, AO, DPC) hacked
%Control: key (0)
%Control: author (72) initials jnrlst
%Control: editor formatted (1) identically to author
%Control: production of article title (-1) disabled
%Control: page (0) single
%Control: year (1) truncated
%Control: production of eprint (0) enabled
%
 
\end{document}